\documentclass{aa}  

\usepackage{graphicx}
\usepackage{txfonts}
\usepackage{hyperref}
\usepackage{color}
\usepackage{pdflscape}
\usepackage{multirow}

\usepackage{hyperref}
\hypersetup{colorlinks,linkcolor=red,citecolor=blue,urlcolor=blue}

\DeclareRobustCommand{\ion}[2]{\textup{#1\,\textsc{\lowercase{#2}}}}
\DeclareRobustCommand{\teff}{T_{\mathrm{eff}}}
\DeclareRobustCommand{\logg}{\log g}
\DeclareRobustCommand{\mh}{\mathrm{[M/H]}}
\DeclareRobustCommand{\feh}{\mathrm{[Fe/H]}}
\DeclareRobustCommand{\vmic}{\varv_\mathrm{mic}}

\DeclareRobustCommand{\ispec}{\texttt{iSpec} }
\DeclareRobustCommand{\kms}{\mathrm{km\,s}^{-1}}
\DeclareRobustCommand{\vr}{v_\mathrm{r}}
\DeclareRobustCommand{\rgc}{R_{\mathrm{GC}}}
\DeclareRobustCommand{\dexGyr}{\mathrm{dex\,Gyr^{-1}}}

\newcommand{\NClusters}{47}
\newcommand{\NGoodstars}{209}
\newcommand{\NRepstars}{33}
\newcommand{\NElements}{25}

\newcommand{\NStarsOutV}{8}
\newcommand{\NStarsOutAP}{34}

\newcommand{\NOCOCCASO}{39}
\newcommand{\NSpecOCCASOi}{236}
\newcommand{\NStarsOCCASOi}{219}

\begin{document} 

  \title{Abundance-age relations with red clump stars in open clusters\thanks{Thanks to observations at Telescope Bernard Lyot, Mercator Telescope and Nordic Optical Telescope, and data retrieved from the archives: ESO, FIES, ELODIE, ESPaDOnS and NARVAL}}

  \author{L. Casamiquela\inst{1} \and C. Soubiran\inst{1} \and P. Jofr\'e\inst{2} \and C. Chiappini\inst{3} \and N. Lagarde\inst{4} \and Y. Tarricq\inst{1} \and R. Carrera\inst{5} \and C. Jordi\inst{6} \and L. Balaguer-N\'u\~nez\inst{6} \and J. Carbajo-Hijarrubia\inst{6} \and S. Blanco-Cuaresma\inst{7}}

  \institute{Laboratoire d’Astrophysique de Bordeaux, Univ. Bordeaux, CNRS, B18N,  allée Geoffroy Saint-Hilaire, 33615 Pessac, France\\
    \email{laia.casamiquela-floriach@u-bordeaux.fr}
  \and
  N\'ucleo de Astronom\'ia, Facultad de Ingenier\'ia y Ciencias, Universidad Diego Portales, Av. Ej\'ercito 441, Santiago, Chile
  \and
  Leibniz-Institut f\"ur Astrophysik Potsdam (AIP), An der Sternwarte 16, 14482 Potsdam, Germany
  \and
  Institut UTINAM, CNRS UMR6213, Univ. Bourgogne Franche-Comté, OSU-THETA Franche-Comté-Bourgogne, Observatoire de Besançon, BP 1615, 25010 Besançon Cedex, France
  \and
  INAF-Osservatorio Astronomico di Padova, vicolo dell’Osservatorio 5, 35122 Padova, Italy
  \and
  Dept. FQA, Institut de Ciències del Cosmos (ICCUB), Universitat de Barcelona (IEEC-UB), Martí Franquès 1, E08028 Barcelona, Spain 
  \and
  Harvard-Smithsonian Center for Astrophysics, 60 Garden Street, Cambridge, MA 02138, USA
  }

  \date{Received ; accepted }

  \abstract
   {Precise chemical abundances coupled with reliable ages are key ingredients to understand the chemical history of our Galaxy. Open Clusters (OCs) are useful for this purpose because they provide ages with good precision.}
   {The aim of this work is to investigate the relations of different chemical abundance ratios vs age traced by red clump (RC) stars in OCs.}
   {We analyze a large sample of \NGoodstars\ reliable members in \NClusters\ OCs with available high-resolution spectroscopy. We applied a differential line-by-line analysis to provide a comprehensive chemical study of \NElements\ chemical species. This sample is among the largest samples of OCs homogeneously characterized in terms of atmospheric parameters, detailed chemistry, and ages.}
   {In our metallicity range ($-0.2<\mh<+0.2$) we find that while most Fe-peak and $\alpha$ elements have flat dependence with age, the $s$-process elements show decreasing trends with increasing age with a remarkable knee at 1 Gyr. For Ba, Ce, Y, Mo and Zr we find a plateau at young ages (< 1 Gyr). We investigate the relations of all possible combinations among the computed chemical species with age. We find 19 combinations with significant slopes, including [Y/Mg] and [Y/Al]. The ratio [Ba/$\alpha$] is the one with the most significant correlations found.}
   {We find that the [Y/Mg] relation found in the literature using Solar twins is compatible with the one found here in the Solar neighbourhood. The age-abundance relations show larger scatter for clusters at large distances ($d>1$ kpc) than for the Solar neighbourhood, particularly in the outer disk. We conclude that these relations need to be understood also in terms of the complexity of the chemical space introduced by the Galactic dynamics, on top of pure nucleosynthetic arguments, especially out of the local bubble.}

   \keywords{ (Galaxy:) open clusters and associations: general--
               Stars: abundances--
               Techniques: spectroscopic }

   \maketitle

\section{Introduction}

To advance towards a broader understanding of the chemical evolution of the Milky Way, detailed element abundances from high-quality spectroscopic data and precise ages are needed \citep{Soderblom2010,Jofre+2019}.
Stars change their chemical composition due to several nucleosynthesis processes, and particularly during their deaths, which lead to an evolution of the overall chemical composition of galaxies. Newly synthesized elements will be inherited to form the next generation of stars, which ultimately guides the chemical evolution in the Universe \cite[see][for reviews]{Freeman+2002, Matteucci2012}. Therefore, the dependency of different element abundances with time is one of the most informative constraints to shape chemical evolution models.

It is usually assumed that the measured atmospheric chemical abundances are representative of the chemical pattern of the interstellar medium at the stars' birthplaces. However, there are some processes happening through a star lifetime which may change their atmospheric abundances.
For instance, in FGK low-mass stars, internal transport processes (e.g. rotation-induced mixing, convection, atomic diffusion, internal gravity waves, or thermohaline mixing) can alter surface abundances of some chemical species such as Li and Be \citep[see e.g.][]{Smiljanic+2016,Lagarde+2017,Charbonnel+2020}. However, the overall consensus is that broadly speaking, chemical abundances of heavier elements measured in the spectra of low-mass stars remain constant during their lifetime, and reflect the chemical composition of their birth clouds.   

\vspace{0.5 cm}

Studying the temporal evolution of chemical abundances of reliable tracers at different regimes of age, metallicity, and Galactic location is crucial to understand the variables that control Galactic evolution, such as supernovae yields, the star formation rate, or feedback mechanisms \citep{Matteucci2012, Maiolino+2019}. The fact that not all chemical elements are produced in the same way or in the same timescale across cosmic times implies we can use a variety of chemical abundance ratios to disentangle the aforementioned different processes affecting galaxy evolution. Hence, it has been proven useful to classify the chemical abundances in families which reflect their main production sites and nucleosynthetic channels: iron-peak, $\alpha$-elements, light elements, odd-Z, and neutron-capture elements. So far, the abundance ratio relating the $\alpha$-capture with the iron-peak family  ($[\alpha\mathrm{/Fe}]$) has been widely considered as a proxy for ageing stellar populations in different Galactic studies \citep[e.g.][]{Minchev+2013}.
However, it is important to keep in mind that there is a scatter in the $[\alpha\mathrm{/Fe}]$ vs. age relation in any location of the Milky Way that can be attributed to the mixture of stellar populations \citep{Chiappini+2015,Anders+2018,Miglio+2020}.

Some other abundance ratios of elements coming from different families could be more informative for chemical evolution investigations because they can have even stronger dependences on stellar age than the $[\alpha\mathrm{/Fe}]$ ratio. These abundance ratios have been recently dubbed as "chemical clocks" \citep[][]{Nissen2015}. 
Indeed, \citet{DaSilva+2012}, and later \citet{Nissen2015}, performed very high precision abundance analysis of solar twins finding a linear trend of [Y/Mg] vs age with slope $\sim 0.04\,\mathrm{dex\,Gyr}^{-1}$. This slope is interpreted invoking the timescale of production of both elements: $\alpha-$ elements (like Mg) are produced by massive stars in a short timescale, while neutron-capture elements produced by the \emph{s}-process (like Y) are mainly synthesized by intermediate-mass stars. The different timescales imply that as time goes on, [Y/Mg] changes significantly more than other abundance ratios. 

This relation has been further confirmed by \citet{TucciMaia+2016} and \citet{Spina+2018} with larger samples of solar twins also analyzed with very high precision. However, when studied in solar-type stars as a function of metallicity \citep{Feltzing+2017}, the relation vanished towards lower metallicities, placing a lower limit of $\feh\sim-0.5$ for the relation to become flat. Later, \citet{Delgado-Mena+2019}, \citet{Titarenko+2019} and \citet{Casali+2020} also noticed a possible non-universality of most chemical clocks at different metallicities, spatial location, and different Galactic populations (e.g. the thick disk). Recently, \citet{Magrini+2021} proposed a theoretical explanation involving magnetic mixing.

To test the behaviour of chemical clocks, it is important to study stars with equivalently precise ages and abundances, but outside the Local Bubble.
An interesting approach to reach a large spatial volume with precise abundances is by analyzing giant stars, but reliable ages are in general difficult to obtain, and particularly at large distances \citep[see][for a review]{Soderblom2010}.
The most reliable methods to obtain ages are those based on stellar evolution models, i.e. isochrone placing or asteroseismology, which presumably are those that depend on the fewest assumptions. 
For individual field stars, isochrone placement becomes challenging and usually, uncertainties in the atmospheric parameters and degeneration of the models prevent obtaining precise ages. Fortunately, placing isochrones is usually much easier in clusters, since they have many coeval mono-metallic members so the results are very precise, which is why they provide historically the primary benchmarks we use to study age-related properties \citep[see][for a review]{Friel2013}.
Asteroseismic ages can be complementary to isochrone ages, as they can be precise \citep[$\sim$30\%][]{Valentini+2019, Miglio+2020} for giants, where isochrone ages cannot give a precise age given the proximity of the evolutionary tracks in the red giant branch.

Open Clusters (OCs) provide an interesting alternative to solar twins to test chemical clocks because their age and distance determinations are much more accurate and precise than for field stars at the same distances. Also, the fact that OCs are chemically homogeneous, at least up to $\sim0.02$ dex \citep[][]{Liu+2016,Casamiquela+2020}, allows for a much more precise chemical abundance determination when several members are observed per cluster, as we shall further see in this paper.
That is why they have been used extensively to trace the Galactic thin disc abundance patterns \citep[e.g.][]{Magrini+2017,Donor+2018,Casamiquela+2019}.

\citet{Slumstrup+2017} showed that the [Y/Mg] vs age relation obtained in the literature, seems to be also followed by red clump giants in 4 OCs, although they used one star per cluster. 
Recently, \citet[][hereafter Paper I]{Casamiquela+2020} complemented this study by analyzing several red giants from other 3 clusters. These clusters also followed well the trend of [Y/Mg] vs age reported in the literature.
However, when \citet{Casali+2020} calculated "chemical ages" from their relation obtained from their 500 solar analogues on a sample of 19 OCs from the \emph{Gaia}-ESO Survey, and compared these ages with those obtained from isochrone fitting, they did not find a robust agreement for all clusters. They concluded that the chemical clocks broke for some of their innermost clusters ($\rgc<7$ kpc), and thus pointing towards a non-universality in the [s/$\alpha$]-[Fe/H]-age relationship.

After \emph{Gaia} DR2 \citep{GaiaCollaboration+2018} new reliable memberships of previously known and newly discovered OCs have been reported \citep[e.g.][]{Cantat-gaudin+2018,Castro-Ginard+2018,Castro-Ginard+2019,Castro-Ginard+2020}. \citet[][CG20 hereafter]{CantatGaudin+2020} have derived ages using an automatic data-driven approach from \emph{Gaia} photometry and parallax for 1867 OCs. 
This sample offers a perfect opportunity to study the trends of chemical abundances and ages for a large number of OCs covering a much larger area of the Galaxy than what reached with solar analogue field stars. In this paper, we present our findings of combining this latest set of OCs with precise ages, with high precision differential abundances obtained from high-resolution spectral analyses.

We organize the paper as follows. In Sect.~\ref{sec:star_sel} we detail the sample of clusters, their membership selections, and the origin of the spectroscopic data, the pipeline used to perform the spectroscopic analysis, and details of the atmospheric parameters. In Sect.~\ref{sec:diffchemab} we describe the computation of the differential chemical abundances and their uncertainties. The OC ages are discussed in Sect.~\ref{sec:ages}.
In Sect.~\ref{sec:XFe_age} we present the cluster chemical abundances [X/Fe] as a function of age. 
In Sect.~\ref{sec:chemclocks} we analyse the possible combinations of abundance ratios which give a significant linear correlation with age. We discuss the obtained linear relations of the [Y/Mg] and [Y/Al] chemical clocks compared with the literature. In Sect.~\ref{sec:beyond} we discuss the effects of using an extended sample of clusters outside of the Solar neighbourhood in the abundance-age relations.
Finally, our conclusions are presented in Sect.~\ref{sec:conclusions}.

\section{The spectroscopic sample}\label{sec:star_sel}

\subsection{Target selection and spectra}
In this study, we have selected the clusters from the revisited list of CG20, which includes ages, distances and extinctions (see an in-depth discussion of the ages in Sect.~\ref{sec:ages}).

We have done a selection of nearby clusters ($d\sim500$ pc): for being closeby, these clusters tend to be well detached from the field distribution in proper motion/parallax, and their membership selection is usually very reliable. We have used membership selections of OCs based on \emph{Gaia} DR2 provided by \citet{GaiaCollaborationB+2018} and by \citet{Cantat-gaudin+2018}, and we have selected the targets according to their positions in the colour-magnitude diagram (in a wide region around the red clump).
The selected stars are usually bright so we could find archival data. We searched for high-resolution and high signal-to-noise ratio (S/N) spectra of the stars from public archives in eight different spectrographs: UVES@VLT, FEROS@2.2m MPG/ESO, HARPS@3.6m ESO, HARPS-N@TNG, FIES@NOT, ESPaDOnS@CFHT, NARVAL@TBL, ELODIE@1.93m OHP (more details of the used instruments are explained in Paper I). We collected spectra corresponding to 151 stars in 11 clusters.
Additionally, we have performed our own observing programs with the spectrograph NARVAL@TBL during 2 semesters (2018B, 2019A). We observed 31 stars in four different clusters. These spectra were reduced using the standard reduction pipeline in NARVAL.

We have added to our sample the spectra observed in the framework of the OCCASO survey \citep{Casamiquela+2016,Casamiquela+2019}, from which the OCs spread a wide range of age and Galactocentric radius. In this case, we selected the highest S/N spectra observed by OCCASO, from the spectrographs HERMES@Mercator and FIES@NOT (Observatorio del Roque de los Muchachos). Before the analysis we already rejected the stars identified as non-members or spectroscopic binaries in the subsample of OCCASO analyzed by \citep{Casamiquela+2019}. Our analysis pipeline succeeded for \NSpecOCCASOi\ spectra in this subsample corresponding to \NStarsOCCASOi\ stars in \NOCOCCASO\ OCs. Several clusters in this sample also had stars with spectra in public archives.

\subsection{Radial velocities and atmospheric parameters}\label{sec:pipeline}
To perform the analysis of the spectra we used the public spectroscopic software \ispec \citep{BlancoCuaresma+2014,BlancoCuaresma2019}. We make use of the same pipeline built in Paper I (we refer the reader to that paper for further details). In brief, the strategy goes as follows:

\begin{itemize}
 \item As a first step, the spectra coming from the different instruments are homogenised cutting them to a common wavelength range (480-680 nm), and downgraded to a common spectral resolution ($45,000$). After the determination of radial velocities using a cross-correlation algorithm, telluric regions and emission lines due to cosmics are masked, and a normalization is performed using a median and maximum filtering.

 \item Afterwards, we use the spectral synthesis fitting method to compute atmospheric parameters ($\teff$, $\logg$, $\mh$, $[\alpha\mathrm{/M}]$, and $\vmic$). The fitting is done by comparing the uncertainty weighted fluxes of a set of observed features with a synthetic spectrum. Atmospheric parameters are varied until convergence of a least-squares algorithm is reached. All the process made use of the LTE\footnote{Local Thermodynamic Equilibrium} radiative transfer code SPECTRUM \citep{Gray+1994} and the MARCS atmospheric models \citep{Gustafsson2008}. We used the \emph{Gaia}-ESO line list \citep{Heiter+2015b}.
\end{itemize}

\begin{figure}
\centering
\includegraphics[width=0.5\textwidth]{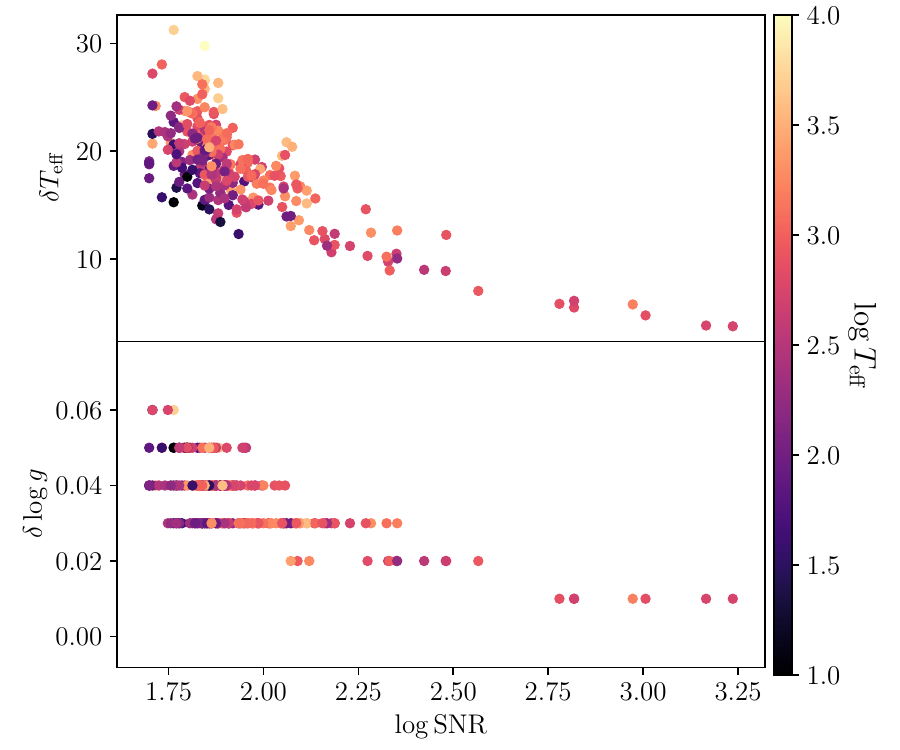}
\caption{Uncertainties in the retrieved $\teff$ (top) and $\logg$ (bottom) as a function of the S/N. The colors show the computed $\log \teff$. }\label{fig:AP_unc}
\end{figure}

Individual uncertainties in the radial velocities, atmospheric parameters and abundances are computed by our pipeline.
We show in Fig.~\ref{fig:AP_unc} the dependence of the uncertainties in the atmospheric parameters as a function of the S/N. Their mean values are 0.045 dex in $\logg$, and 18 K in $\teff$. Uncertainties are also slightly dependent on the temperature of the star: colder stars exhibit lower uncertainties at a given S/N. 

\subsubsection*{Zeropoints among instruments}
To account for systematic offsets between the different instruments, we analyzed the computed atmospheric parameters of stars observed with several instruments. In Paper~I, we did the same exercise and we found no systematic difference above the quoted uncertainties. In the present work, we have a different sample of stars, composed only by red giants (mostly RC), and we have added an additional instrument used for the OCCASO observations, HERMES. Therefore, we recompute the offsets and dispersions for this new sample with the \NRepstars\ stars observed with different instruments. The results are listed in Table~\ref{tab:diff_AP}. We show, for each instrument, the median difference in $\teff$, $\logg$ and $\mh$ (overall metallicity) of stars in common with that given instrument and all of the others. In general, offsets are consistent with the dispersions with very few exceptions, the largest one being the offset of $0.05\pm0.02$ dex in $\logg$ found with NARVAL with only 2 stars in common with other instruments. For this same instrument, we found a smaller offset in Paper I, using a larger sample of stars (including main sequence stars): $0.02\pm0.02$ dex.

In our range of atmospheric parameters, we do not find any star in common with other instruments, either for HARPSN and ELODIE. In Paper I we found small offsets with ELODIE: $2\pm21$ K in $\teff$, $-0.05\pm0.04$ dex in $\logg$ for main sequence stars.

The obtained comparisons are fully consistent with the mean uncertainties quoted by our pipeline and shown in Fig.~\ref{fig:AP_unc}.

\begin{table}
\caption{Median offsets and dispersions (MAD) found in $\teff$, $\logg$ and $\mh$ among stars observed with several instruments. The number of measurements used is indicated in the last column.}\label{tab:diff_AP}
\centering
\setlength\tabcolsep{3pt}
\begin{tabular}{lrrrr}
 \hline
Instrument  & $\Delta\teff$ [K] & $\Delta\logg$ & $\Delta\mh$ & N  \\
\hline
FEROS    & $ 8 \pm 14 $ & $ 0.02 \pm 0.06$ & $ 0.02 \pm 0.01$ & 11 \\
ESPADONS & $ 13 \pm 10$ & $ 0.02 \pm 0.04$ & $ 0.02 \pm 0.03$ &  2 \\
HARPS    & $  1 \pm 20$ & $-0.03 \pm 0.04$ & $-0.01 \pm 0.01$ & 22 \\
NARVAL   & $ 15 \pm 20$ & $ 0.05 \pm 0.02$ & $ 0.03 \pm 0.04$ &  2 \\
UVES     & $ -3 \pm 24$ & $ 0.02 \pm 0.04$ & $-0.01 \pm 0.04$ & 13 \\
FIES     & $ -4 \pm 29$ & $ 0.01 \pm 0.04$ & $ 0.01 \pm 0.03$ & 17 \\
HERMES   & $-18 \pm 16$ & $-0.04 \pm 0.04$ & $-0.02 \pm 0.01$ & 21 \\
\hline
\end{tabular}
\end{table}

\subsection{Membership refinement and red clump selection}
As in Paper I, we have done a membership refinement using the radial velocities computed with iSpec, to reject stars which have 3D kinematics not consistent with that of the cluster, which are candidates to be non-members or spectroscopic binaries.
For the three clusters already analyzed in Paper I (Hyades, NGC~2632 and Ruprecht~147) we kept the same selection of member stars.
For the other clusters, we calculate Galactic 3D velocities $(U,V,W)$ from proper motions, parallax and positions from \emph{Gaia} DR2, and our own radial velocities.
Then, we compare 3D velocity of our stars with the distribution of the cluster stars given by the secure members \citep[][$p_{memb}>0.7$]{Cantat-gaudin+2018} with radial velocities in \emph{Gaia} RVS. We manually discard as members those stars in our spectroscopic sample too discrepant in the $(U,V,W)$ distribution of  each cluster (see Fig~\ref{fig:exampleUVW} as an example). Even though the precision in the \emph{Gaia} radial velocities is much lower than that coming from iSpec, it is straightforward to identify the most discrepant cases if we also take into account the internal consistency among the spectroscopic targets. We rejected \NStarsOutV\ stars.

We realized that some of the remaining stars appear to have $\teff$ and $\logg$ away from the red clump, tipically stars in the asymptotic giant branch (AGB) or in the subgiant branch (see Fig~\ref{fig:exampleHR} as an example). We manually rejected those stars, because for the purpose of this paper we require that the atmospheric parameters of the stars to be as similar as possible (see details of the method in Sect.~\ref{sec:diffchemab}). This procedure is easier for the clusters with several stars observed, because they get visually clumped in the Kiel diagram, and is less reliable for the clusters where only one or two stars were observed. We rejected a total of \NStarsOutAP\ stars with this criterion.

\begin{figure}
\centering
\includegraphics[width=0.5\textwidth]{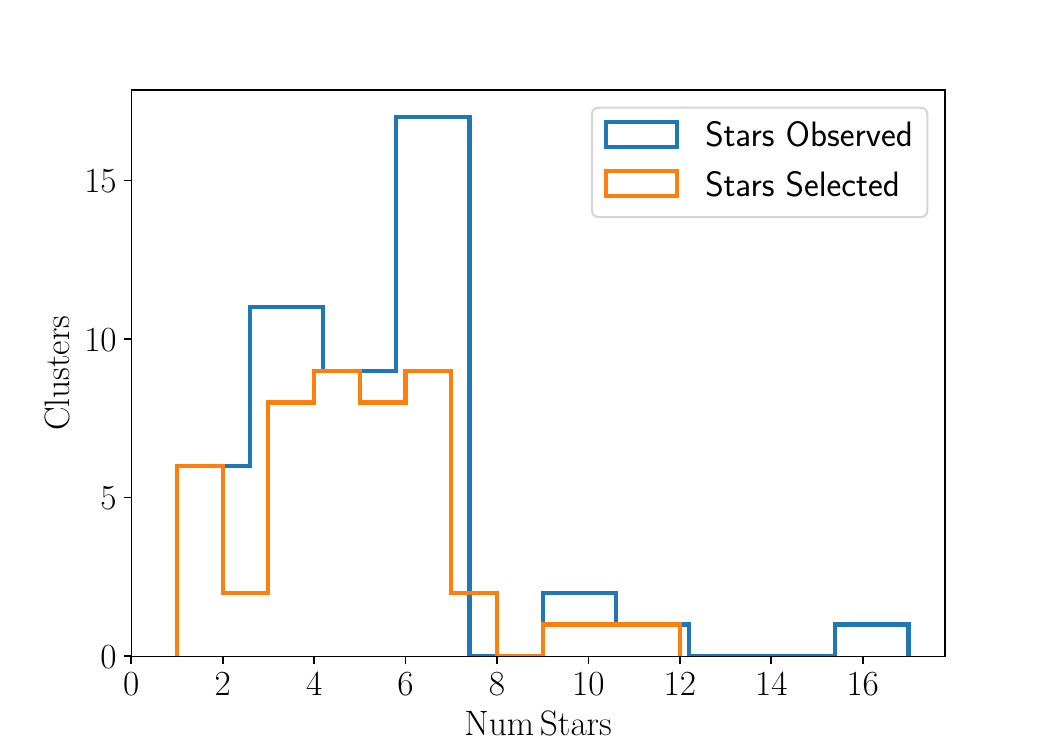}
\caption{Number of clusters which have a given number of observed spectroscopic targets (blue), and those considered in our analysis as bona fide red clump stars.}\label{fig:hist_Nstars}
\end{figure}

In the end, we consider as reliable bona fide red clump members \NGoodstars\ stars in \NClusters\ OCs. The details of the spectra of these stars are in Table ~\ref{tab:stars}. The S/N of the used spectra ranges between 50 and 1780. The clusters have between 1 and 12 stars per cluster, as plotted in Fig.~\ref{fig:hist_Nstars}. The properties of the clusters are listed in Table~\ref{tab:clusters}.

\subsection{The final sample}\label{sec:FinalSample}

\begin{figure}
\centering
\includegraphics[width=0.5\textwidth]{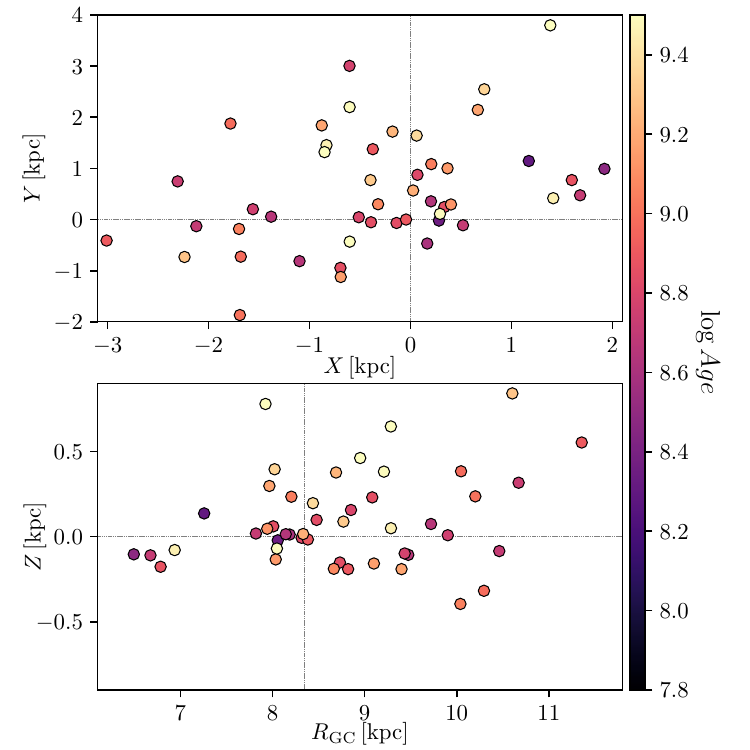}
\caption{Top: $X$,$Y$ distribution of the sample of OCs used in this work, where the Galactic center is towards the right. Bottom: $Z$, $\rgc$ distribution of the sample of clusters. The color represents the age of the cluster. The dashed lines indicate the position of the Sun $(X_{\odot},Y_{\odot},Z_{\odot})=(0,0,0)$, $\rgc=8.34$ kpc.}\label{fig:XY}
\end{figure}

\begin{figure*}
\centering
\includegraphics[width=\textwidth]{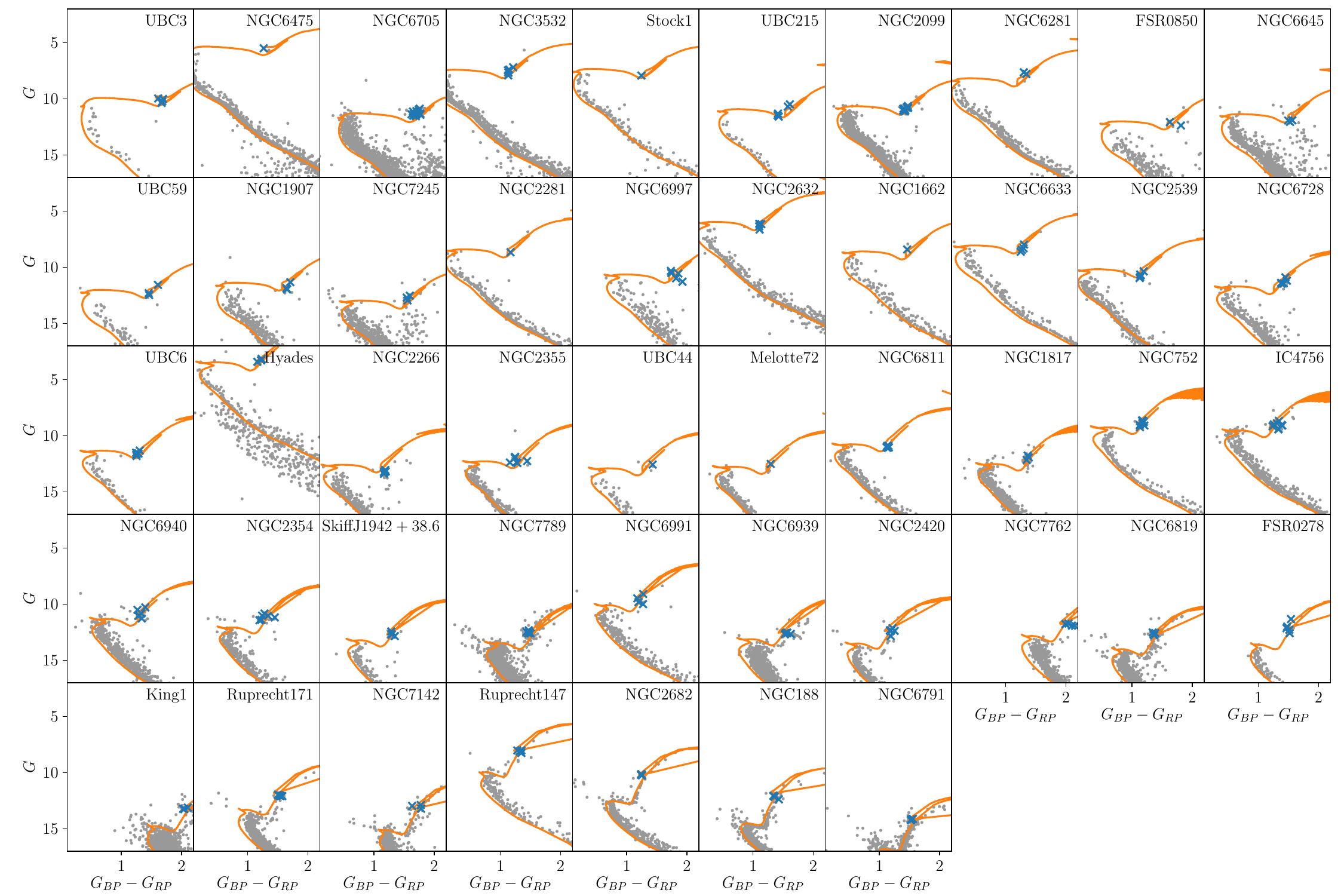}
\caption{Gaia DR2 color-magnitude diagrams of the sample of clusters used (sorted by age), retained RC targets are marked with a blue cross. The grey points correspond to the membership selections by \citet{Cantat-gaudin+2018,GaiaCollaborationB+2018}. PARSEC isochrones \citep{Marigo+2017} of +the corresponding metallicity computed in this paper are overplotted using distances provided by CG20, and ages/extinctions detailed in Sect.~\ref{sec:ages}.}\label{fig:HR}
\end{figure*}

We plot in Fig.~\ref{fig:XY} the $(X,Y)$ and $(Z,\rgc)$ distribution of the analyzed OCs. Our entire sample includes clusters preferentially in the first, second and third Galactic quadrants. The clusters are confined in the thin disk, and only a few old clusters go beyond $Z=500$ pc. Contrary to most of the samples where chemical clocks are analyzed, half of our sample is outside the 1 kpc local bubble around the Sun. In Fig.~\ref{fig:HR} we plot the colour-magnitude diagram (CMD) of all clusters using the available memberships, with the retained RC stars. A table with a complete list of clusters including their ages, Galactic positions, number of stars observed, number of spectra and mean abundance results is provided in Tab.~\ref{tab:clusters}.

Our sample includes 5 clusters which were discovered after \emph{Gaia} DR2 \citep[UBC clusters][]{Castro-Ginard+2020}, therefore this is the first time that chemical abundances are determined for them. The details of these results will be found in a paper in preparation.

\begin{figure*}
\centering
\includegraphics[width=\textwidth]{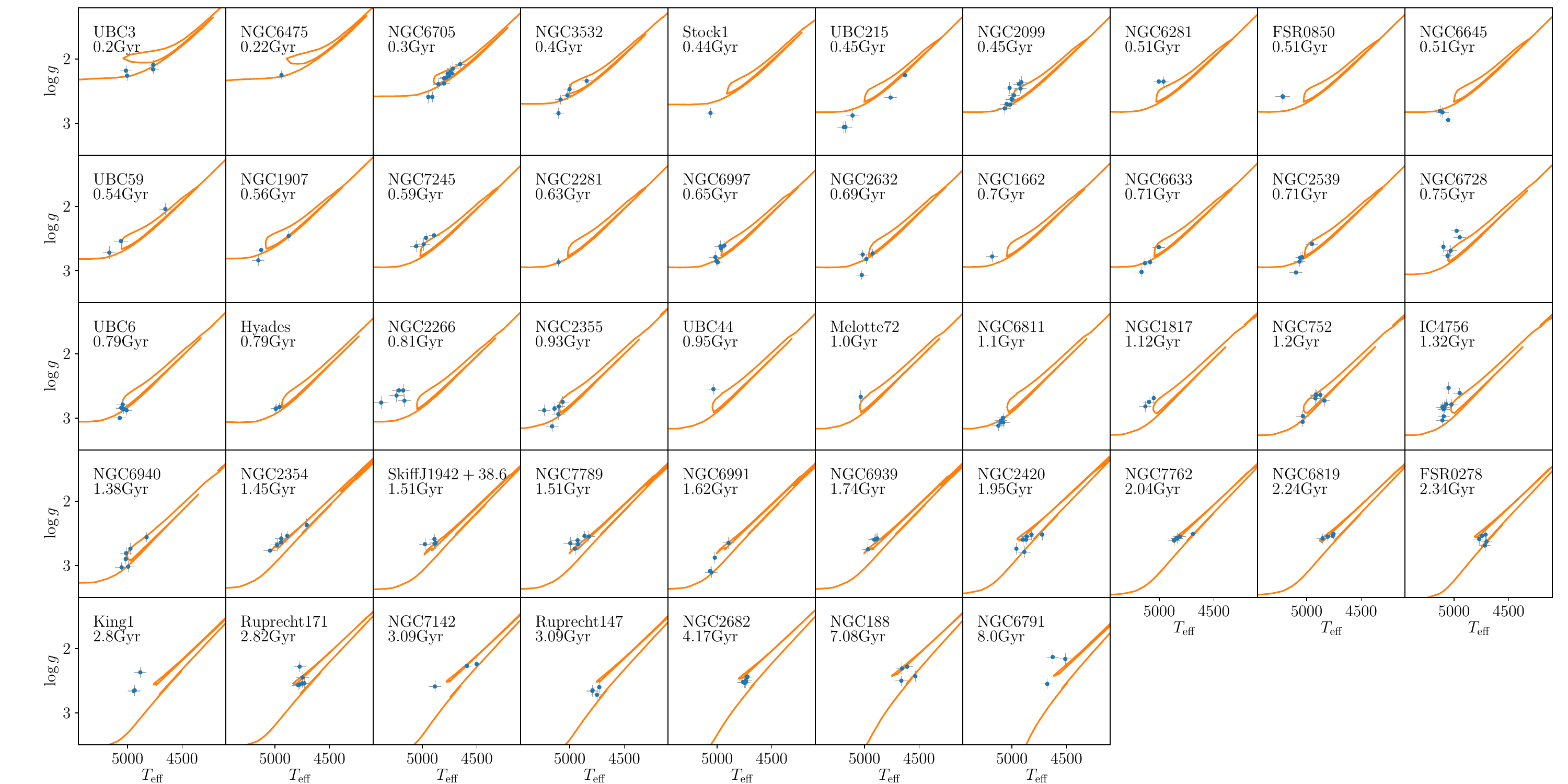}
\caption{$\teff$-$\logg$ diagrams resulting from the analysis of the analyzed RC stars in blue. The clusters are sorted according to age. The same isochrones of Fig.~\ref{fig:HR} are overplotted with the corresponding [M/H] and age.}\label{fig:AP}
\end{figure*}

The results of $\teff$ and $\logg$ per star and cluster are shown in Fig.~\ref{fig:AP}. The clusters are sorted by increasing age as determined in Sect.~\ref{sec:ages}, and we overplot the corresponding isochrone of the [M/H] determined in this work, to guide the eye. We see that in almost all clusters the isochrone corresponds well to the parameters derived for the member RC stars. There are few cases where the retrieved atmospheric parameters are off from the expected RC traced by the isochrone. We do not find any relation of this with the age or any atmospheric parameter, but this usually happens for some stars at the low S/N limit. 
In the case of King~1, we highlight that it is among the faintest cluster in the sample, has a mean $\feh=0.3$, and its CMD in Fig.~\ref{fig:HR} shows that its membership selection can be uncertain. For the clusters where the mean of computed $\teff$ and $\logg$ values are larger by at least $150$ K or $0.2$ dex w.r.t. the expected RC position of the model (Stock~1, NGC~6281, FSR~0850,  NGC~2266, UBC~44, Melotte~72 and King~1), we keep an internal flag to discuss any possible effect that they may introduce in the abundance-age trends.

\section{Differential chemical abundances}\label{sec:diffchemab}
Our aim is to compute chemical abundances of \NElements\ species in a line-by-line differential manner for all stars, w.r.t. a reference star. The set of lines to be used for the differential chemical abundance results were selected in the same way as in Paper I.
In a first step, individual absolute chemical abundances per spectrum are obtained line-by-line using the atmospheric parameters fixed to the resulting values of Sect.~\ref{sec:pipeline}. We use the same radiative transfer code (SPECTRUM), model atmospheres (MARCS), line list and fitting algorithm as in Sect.~\ref{sec:pipeline}. Hyperfine structure splitting and isotopic shifts have been taken into account for the elements: V I, Mn I, Co I, Cu I, Ba II, La II, PrII, and Nd II, following \citet{Heiter+2015b,Heiter+2020}.
Our analysis did not take into account neither non-LTE nor 3D corrections in the computation of the abundances. These corrections are in general small for nearly-solar metallicity stars \citep{Amarsi+2020}. In the case of this study, we have a narrow range of atmospheric parameters in our sample of stars and we perform a differential treatment of the abundances. This type of analysis helps to mitigate the possible departures from 1D/LTE, which are highly dependent on the used lines and the stellar parameter range.

The differential abundance analysis provides high-precision abundance measurements, erasing most of the effects that introduce systematic uncertainties in the usual abundance computations (e.g. blends, wrong atomic line parameters). To be able to apply this technique successfully, it is required that the sample of stars is in the same evolutionary stage as the reference star. A clear example of this is the analysis of solar twins with respect to the Sun, largely applied in the literature usually using nearby solar twins \citep[e.g.][]{Melendez+2009b, TucciMaia+2016}, or solar analogues in clusters \citep[e.g.][]{Liu+2016}. Other works have applied the same strategy to other stellar types using a reference star very close to the sample of stars \citep[e.g.][]{Hawkins+2016b,Jofre+2015,BlancoCuaresma+2018}, as we did in Paper I.

All analysed stars here are selected to match the RC position of the cluster, or very close, as it was discussed in Sect.~\ref{sec:star_sel}. However, the RC position slightly moves in the Kiel diagram as age/metallicity changes, in general, reproduced by our $\teff$-$\logg$ results. Therefore, since the cluster sample spans a large range in age, stars do not have exactly the same atmospheric parameters, though are considered to be in the same evolutionary state. We consider the stars similar enough to be able to apply this procedure. Additionally, for each cluster, we have checked that the differences in the atmospheric parameters of the selected stars do not produce trends in abundances, e.g. it does not exist any dependence of abundances vs $\teff$ or $\logg$ (see an example in Fig.~\ref{fig:NGC3532}). Doing this we aim to minimize any possible bias in the abundances caused by differences in the atmospheric parameters of the selected stars.

\subsection{Setting the reference values}
As a reference star we use the giant from the Hyades Gaia DR2 3312052249216467328 (gam Tau/HD 27371), also used as a reference in Paper I. From the analysis of two spectra of this star, we obtained mean atmospheric parameters $\teff=4975 \pm 12$ K, $\logg = 2.83 \pm 0.03$, and mean iron abundance $\feh=0.10 \pm 0.05$ (with respect to the Sun).
The mean absolute differences of the full sample of stars with respect to the reference star are 116 K and 0.21 dex, in $\teff$ and $\logg$ respectively, with the maximum differences being 460 K and 0.9 dex. These values are slightly larger than for the samples analyzed in Paper I, because of the aforementioned reasons.
We consider the sample of giants as analogues, close enough to the reference star, to safely perform this analysis. This is further justified since we do not find any remarkable trend of abundances with $\teff$ nor $\logg$, for the cluster stars with the larger range of atmospheric parameters.

The outcome is star-by-star element abundances computed with respect to the abundances of the reference star from the Hyades. Afterwards, we transform the resulting abundances to a solar scale to be able to retrieve meaningful "bracket ratios" of abundances. To do so, we analyze the solar analogue from the Hyades Gaia DR2 3314109916508904064 (HD 28344; $\teff=5957\pm32$ K, $\logg=4.49\pm0.03$, $\feh=0.12\pm0.05$) differentially with respect to the Sun, using the same methodology. Considering the stars within the Hyades to have the same abundance pattern, we can then do the transformation of the abundances of giants to the solar scale, adding to each element abundance the corresponding value of the solar analogue of the Hyades.

\subsection{Final cluster abundance}
\begin{figure}
\centering
\includegraphics[width=0.5\textwidth]{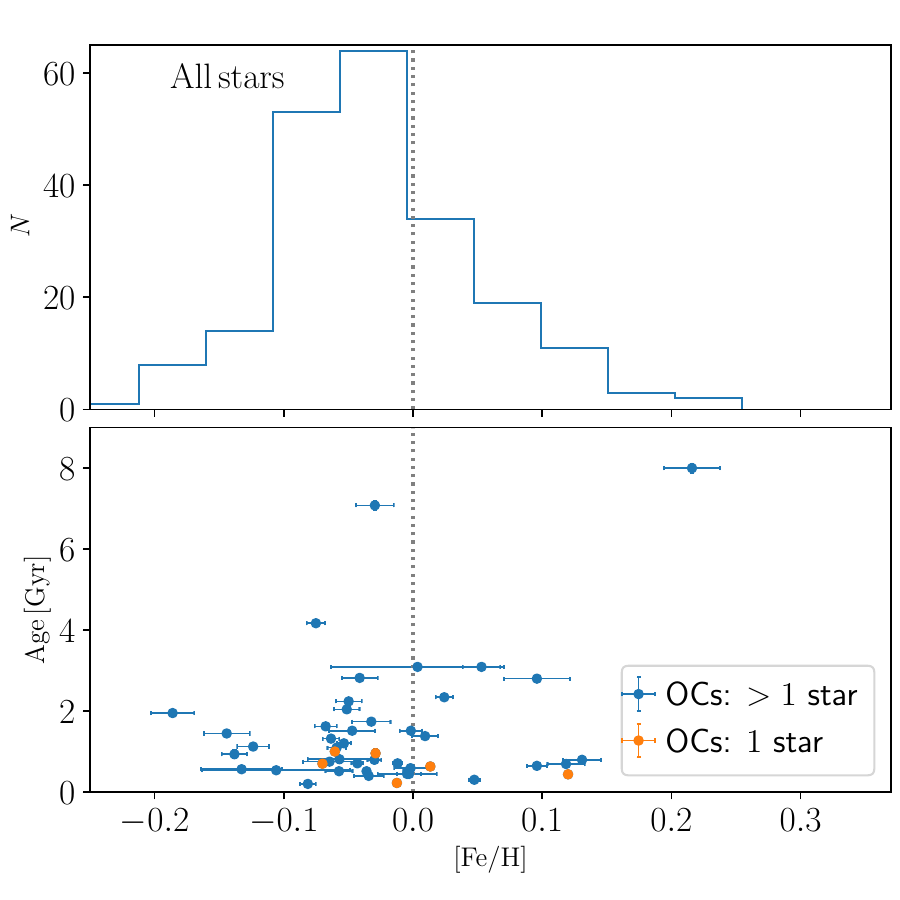}
\caption{Top: distribution of solar scaled iron abundances of all analyzed stars. Bottom: distribution of cluster ages as a function of mean [Fe/H] (clusters with one measured star are plot in orange).}\label{fig:hist_FeH}
\end{figure}

For each cluster, we perform a weighted mean to obtain the cluster mean abundance values, and we use the corresponding dispersion as uncertainties. The resulting cluster abundances are detailed in Tab.~\ref{tab:clusters}. We plot in Fig.~\ref{fig:hist_FeH} the distribution of the retrieved $\feh$ for all used stars, and the distribution of cluster ages as a function of mean $\feh$ per cluster. The sample of clusters is constrained between $-0.2<\feh<0.2$.
Abundances with large uncertainties ($>0.2$ dex, typically corresponding to the most uncertain elements or low S/N spectra) are rejected to compute mean values since we do not consider them reliable for our purpose in this paper. 

\begin{figure}
\centering
\includegraphics[width=0.5\textwidth]{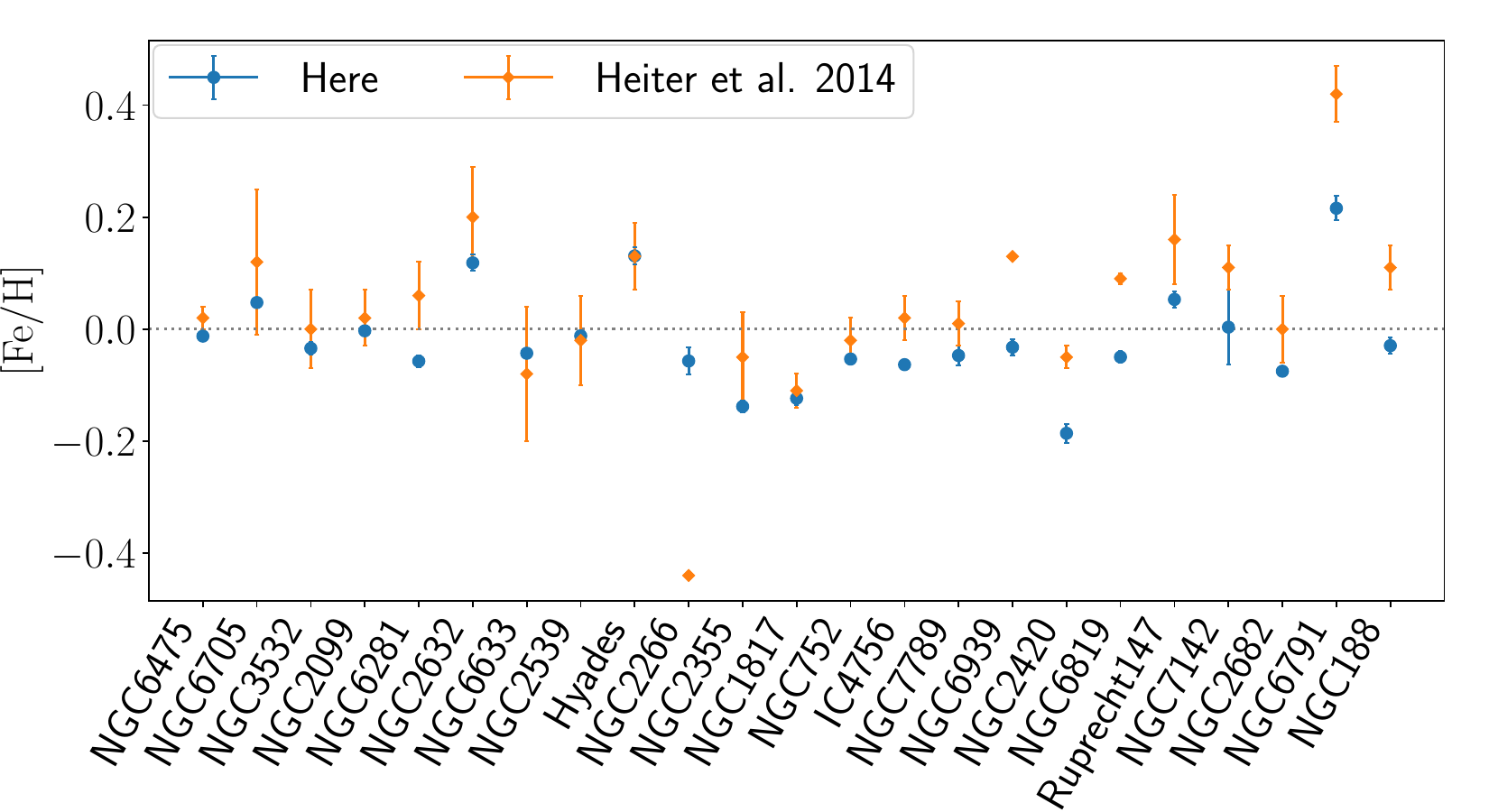}
\caption{Mean [Fe/H] values per cluster computed here (blue), compared with [M/H] literature values from \citet{Heiter+2014} (orange), for the 23 clusters in common.}\label{fig:FeH_literature}
\end{figure}

We show in Fig.~\ref{fig:FeH_literature} the mean [Fe/H] values per cluster retrieved here compared with the catalog of compiled cluster metallicities by \citet{Heiter+2014}. We have 23 clusters in common, and we find a general good agreement ($\sim2\sigma$) with literature values. There are differences in some cases, which are expected due to differences in the quality of the spectra and analysis pipelines.
The most discrepant case is NGC~2266, with a difference of $\sim$0.6 dex, for which the literature value comes from one single star studied in \citet{Reddy+2013}\footnote{The observed star is TYC 1901-558-1,  Gaia DR2 3385733650131889408}. This star is not considered member in \citet{Cantat-gaudin+2018}, because its proper motions and parallax from \emph{Gaia} DR2 differ w.r.t. the mean cluster values. Moreover, our determination for this cluster based in 5 members also points to a different radial velocity: we obtain a mean value of $\vr=52\pm2\,\kms$ while the value of \citet{Reddy+2013} is $-29.7\pm0.2\,\kms$.
NGC~6791 presents a quite large difference of 0.15 dex. It is the most metal-rich cluster in our sample, so possibly uncertainties and systematics in this study and/or literature are larger in this case.

\begin{figure}
\centering
\includegraphics[width=0.5\textwidth]{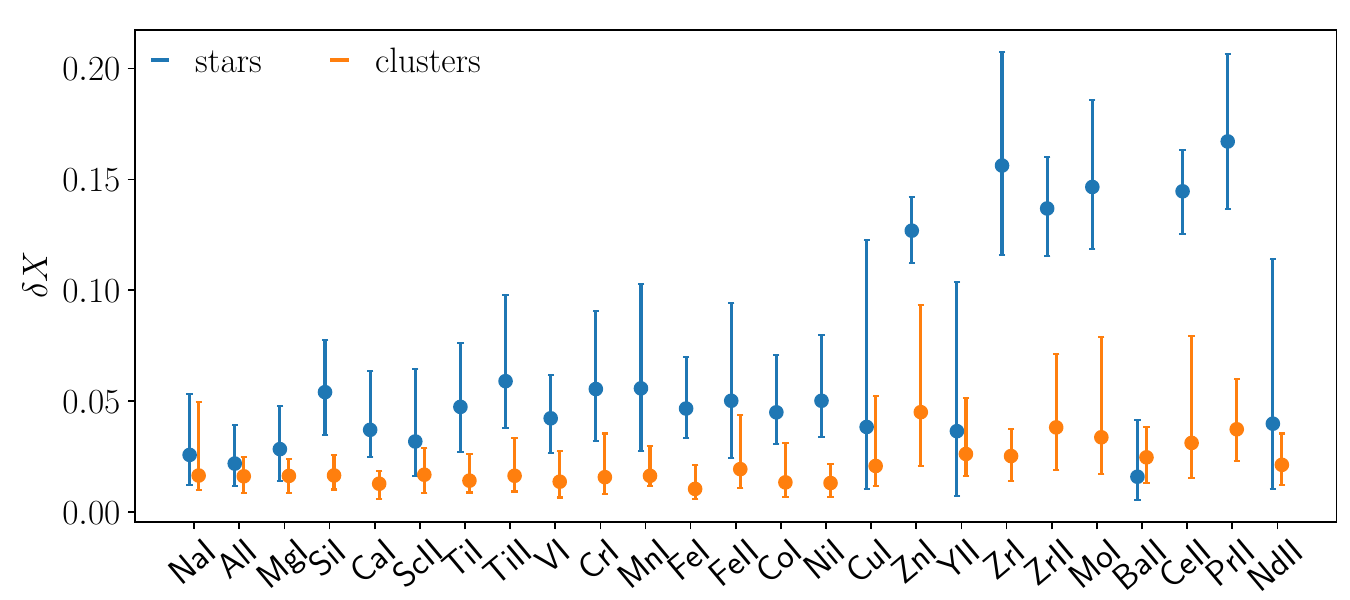}
\caption{Distribution of the uncertainties of abundances of the different elements: for the full sample of stars (blue), and the clusters with more than one measured star (orange). The median value is marked with a point, and the 16th and 84th percentile are represented with the line.}\label{fig:unc_abus}
\end{figure}

We show the distribution of uncertainties in element abundances of individual stars (blue symbols), and in clusters with more than one measured star (orange symbols) in Fig.~\ref{fig:unc_abus}. To compute the cluster individual uncertainties, we used the standard deviation of the member abundances divided by $\sqrt{N_{\mathrm{stars}}}$. Cluster uncertainties are in general much lower than star typical uncertainties, due to the good agreement among the different stars of the same cluster. We highlight that the most uncertain elements in our measurements are Zn, Zr, Mo, Ce and Pr. For these elements, individual stars have typically uncertainties $>0.1$ dex. Cluster uncertainties are found to be the largest also for the same elements ($\sim0.05$ dex), with long tails. These long tails are however driven by few clusters, typically: NGC~6791 and FSR~0850 have the largest uncertainties sometimes reaching 0.3 dex, for Ce, Zn and Mo.

\section{Ages}\label{sec:ages}

\begin{figure*}[ht]
\centering
\includegraphics[width=\textwidth]{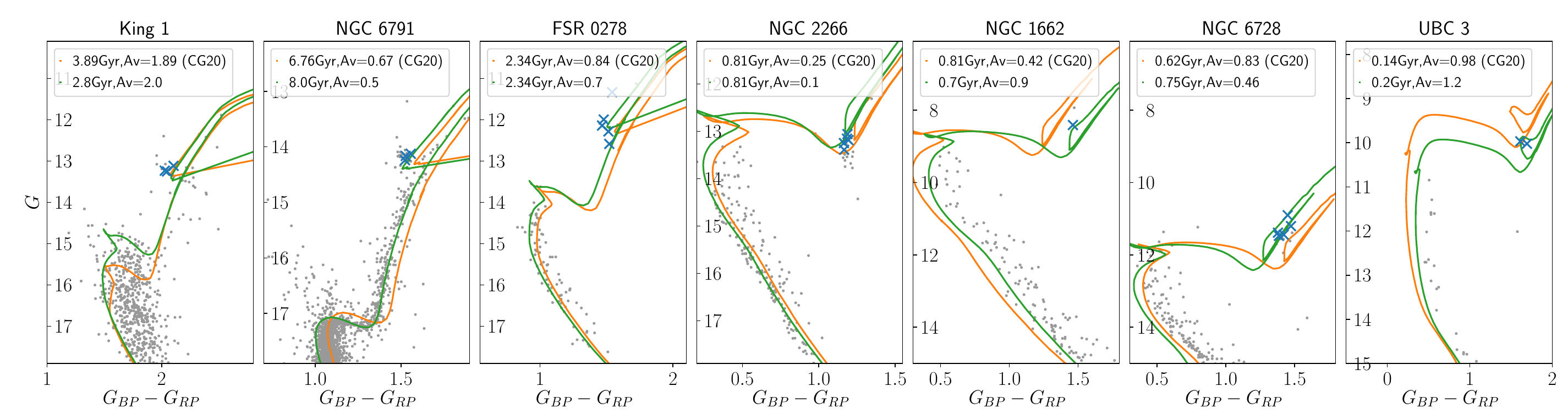}
\caption{\emph{Gaia} DR2 CMDs (gray dots) of the clusters for which the ages and extinctions were refined with respect to the ones in CG20, as discussed in the text. The orange isochrones are those with the values quoted in CG20, and the green ones are manually changed to adjust the shape of the main sequence, turnoff or red giant members. Target stars studied in this paper are marked with blue crosses.}\label{fig:age_change}
\end{figure*}

As starting point for the cluster ages we have used the catalog of ages derived by CG20. That study obtained ages, extinctions and distances for a large set of OCs from an artificial neural network, using \emph{Gaia} photometry and the mean parallax of the clusters. The algorithm was trained using a reference sample composed mainly by \citet{Bossini+2019} (the largest homogeneous sample of cluster ages obtained from \emph{Gaia} data), and complemented with other smaller samples to allow a coverage of older and distant clusters.
CG20 does not provide individual uncertainties per cluster, thus we assign age uncertainties extrapolated from \citet{Bossini+2019}. We took the median of \citet{Bossini+2019} quoted uncertainties per age bin, and used them as estimation of the cluster uncertainty at a given age.

CG20 performed an external comparison of their set of ages, distances, and extinctions with large catalogues such as \citet{Kharchenko+2013}, finding a general good agreement in ages, with exception of some young reddened objects. In the case of our clusters, the quoted ages, extinctions and distance modulus are, in general, visually consistent in a CMD when using PARSEC isochrones. However, CG20 does not take into account particularities of a given cluster, such as metallicity, to assign ages. With a visual inspection of the \emph{Gaia} CMDs we have spotted some cases where the quoted age can be adjusted when using an isochrone of the metallicity derived in Sect~\ref{sec:diffchemab}. In these cases, we have done a manual refinement of the extinction and the age of the cluster, leaving the cluster mean distance as constant, since it is constrained mainly by the mean cluster parallax. We did the following changes (see Fig.~\ref{fig:age_change}):

\begin{itemize}
 \item King~1: the quoted age (3.89 Gyr) and extinction (1.89 mag) by CG20 seem to not fit the turnoff region of the cluster. For its super-solar metallicity of $\mathrm{[Fe/H]}=+0.3$, a younger age of $2.8$ Gyr and a larger absorption of 2 mag present a better fit. This age has a good agreement with previous dedicated studies of this cluster \citep[$2.8\pm0.3$ Gyr][]{Carrera+2017}.
 \item NGC~6791: CG20 quotes an age of 6.76 Gyr and $A_V=0.67$ mag. This cluster also has a super-solar metallicity of $\mathrm{[Fe/H]}=+0.26$. In this case, a decrease of the extinction to $A_V=0.5$ mag and an increase of the age to 8 Gyr, provides a good fit of the turnoff, red giant branch and red clump simultaneusly. Again, in this case the refined age is more compatible with other literature studies such as $7.7\pm0.5$ Gyr \citep{Grundahl+2008} and $\sim8$ Gyr \citep{Brogaard+2012}, both from different sets of isochrones.
 \item For the clusters FSR~0278 ($\mathrm{[Fe/H]}=+0.13$) and NGC~2266 ($\mathrm{[Fe/H]}=+0.15$) the quoted ages seem to reproduce well the shape in the CMD: 2.34 and 0.81 Gyr, respectively. However, a small change in the absorptions, which seems overestimated for both cases in CG20 ($A_V=$0.84, and $A_V=0.25$ mag, respectively) , fits better the red clump and main sequence. We derive $A_V=$0.7 and 0.1 mag, respectively for the two clusters.
 \item The quoted values for NGC~1662 ($\mathrm{[Fe/H]}=-0.06$) in CG20, 0.81 Gyr and $A_V=0.42$ mag, can be slightly refined to 0.7 Gyr and $A_V=$0.9 to fit better the turnoff and red giants.
 \item For NGC~6728 ($\mathrm{[Fe/H]}=0.02$) the quoted parameters in CG20 (620 Myr $A_V=0.83$) fit well the turnoff and main sequence, however the well populated red clump is shifted upwards with respect to the isochrone. A smaller absorption of $A_V=0.46$ and older age of 750 Myr, similar to the one derived by \citet{Bostanci+2018}, fits better the clump, even though compromises slightly the main sequence.
 \item UBC~3 ($\mathrm{[Fe/H]}=-0.05$) values from CG20 (140 Myr and $A_V=0.9$) can be slightly changes to 0.2 Gyr and $A_V=1.2$ mag to provide a more adequate fit of the main sequence and red clump.
\end{itemize}

The ages of our analyzed clusters range between $140$ Myr and $7$ Gyr, most of them having ages $<3$ Gyr.

\begin{figure*}
\centering
\includegraphics[width=\textwidth]{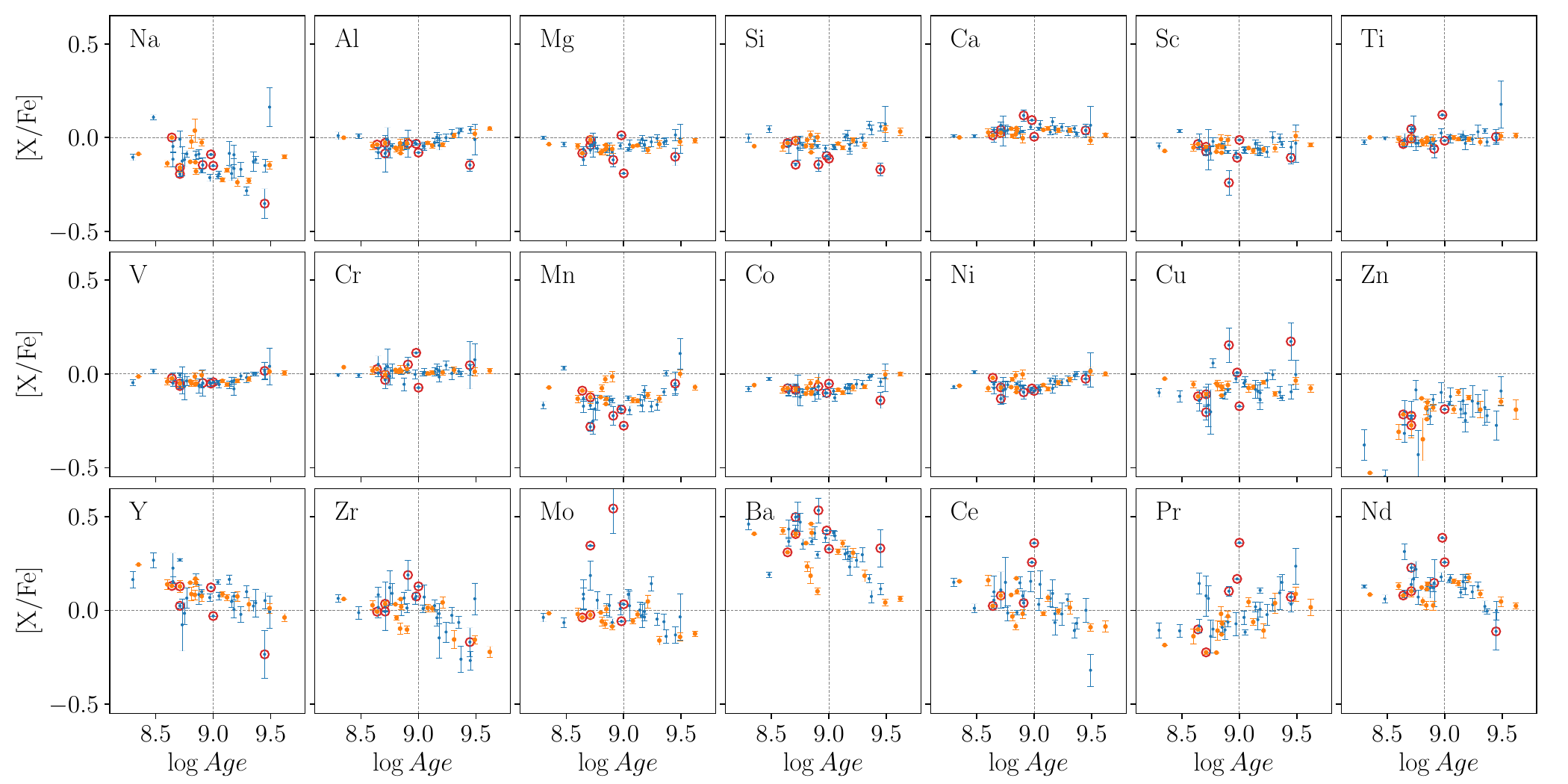}
\caption{[X/Fe] abundances of all measured elements as a function of age in logarithmic scale. Clusters located at less than 1 kpc from the Sun are plot in orange. Uncertainties in x coordinate are smaller than the points. Red circles correspond to the clusters with stars with more uncertain atmospheric parameters.}\label{fig:XFe_age}
\end{figure*}

\section{Chemical trends of [X/Fe] with age}\label{sec:XFe_age}

We plot the [X/Fe]\footnote{From now on, we only use the neutral states of Fe and Ti as representative of element abundances, and the ionized state of Zr. This choice is made after the inspection of the uncertainties of the chemical species.} of all the measured chemical elements as a function of cluster's log age in Fig.~\ref{fig:XFe_age}. We mark the clusters closer than 1 kpc in orange. For clarity, we only plot those clusters for which uncertainties in abundance are lower than 0.15 dex in the specific element. We highlight in red circles the flagged clusters in Sect.~\ref{sec:FinalSample}. A complementary figure with [X/Fe] as a function of [Fe/H] and coloured by age can be found in Fig.~\ref{fig:XFe_FeH}.

Most of the ratios from Na to Ni exhibit a mostly flat behaviour as a function of age, which is expected in the metallicity range of the clusters \citep[see][]{Kobayashi+2020}. However, some of them seem to have a mild increasing trend starting at 1 Gyr (Al, Si, V, Mn, Co, Ni). We remark that Na, Cu and Mn have the largest spreads in these spaces, and a small negative zero point. This can be due to NLTE effects \citep{Yan+2016,Bergemann+2019}, which usually depend on the atmospheric parameters of stars, and that are not taken into account when doing the transformation from differential abundances w.r.t. giants to solar scaled abundances. In several elements (e.g. Mg, Si) the flagged clusters in red tend to be slightly off from the distribution of the other clusters, showing that they probably carry larger uncertainties. Zn also shows a negative zeropoint of the abundances relative to Fe of around -0.2 dex. A similar zeropoint was found by \citet{Duffau+2017} which studied Zn abundances with Gaia-ESO survey data for field giants at solar metallicity. They concluded that this cannot be explained by systematic uncertainties such as NLTE or 3D effects, thus they attributed the underabundances to a real signature due to chemical enrichment. After studying this effect as a function of Galactocentric radius, metallicity and age, they tentatively explained it as due to dilution from almost Zn-free SN Ia ejecta.

The $s$-process elements (Y, Zr, Mo, Ba, Ce, Pr, Nd) are plotted in the third row in Fig.~\ref{fig:XFe_age}.
In general, we observe a change of behaviour of all these elements at $1$ Gyr, traced with less dispersion for the nearby clusters. This is particularly remarkable for Y, Zr, Mo, Ba, Ce and Nd. These elements exhibit a sort of plateau with large scatter at ages younger than $\sim1$ Gyr, but at different "saturation" values, with Ba being the largest one. Older than 1 Gyr, the abundances show a significant negative trend with age. Some of the elements exhibit large dispersions, such as Mo, which makes the decreasing trend less clear. Pr shows the opposite trend, with increasing values as a function of age. Pr has a significant contribution to the $r$-process (see more details in Sect.~\ref{sec:chemclocks}), which could cause this remarkably different behaviour.

Ba is the $s$-process element which shows the largest [X/Fe] abundances for young OCs, which diminishes significantly at older ages ($\sim2$ Gyr) approaching solar values. This is a known matter without a satisfactory explanation yet, investigated by other authors before \citep[e.g.][]{DOrazi+2009,Mishenina+2013,Magrini+2018} with smaller samples of OCs. For the first time \citet{DOrazi+2009} showed a sharp decreasing trend of [Ba/Fe] in a sample of 20 clusters, which could be explained using chemical evolution models with enhanced yields of the $s$-process production by AGB stars below $1.5\mathrm{M_{\odot}}$. Some samples of field stars, for example that of \citet{Bensby+2005}, also show that young stars seem Ba-enhanced, even though their sample is limited to slightly older ages than the OCs, and exhibit a much larger spread in [Ba/Fe]. This is possibly because the ages of field stars computed by isochrone fits are much less accurate than the ages of clusters. \citet{Maiorca+2011} used a similar sample of OCs to that of \citet{DOrazi+2009} and computed other $s$-process elements (Y, Zr, La and Ce) to also find decreasing dependencies. For these four elements they spotted a plateau at ages smaller than $\sim1$ Gyr, but this was not seen for Ba. Later, \citet{Jacobson+2013} argued that not all $s$-process elements follow this decreasing trend, like Zr and La. More recently, \citet{Magrini+2018} analyzed $s$-process elements (Y, Zr, Ba, La, Ce, Eu) in the sample of 22 clusters in the Gaia-ESO survey: they found steep trends in Y, Zr, Ba and lower correlations for La and Ce. They attribute this to the lower number of stars measured for these elements. Special caution needs to be taken in interpreting Ba results, since several studies recently pointed towards Ba II lines being influenced by several effects in young stars. For instance, \citet{Reddy+2017} obtained a strong correlation of stellar activity with the Ba II abundances obtained from the 5853 $\AA$ line. This line forms in the upper layers of the stellar photosphere where the effects of the active chromosphere are strong for young stars ($t<150$ Myr). In our case, we are using also the line at 6141 $\AA$, which gives very consistent abundances with the line at 5853 $\AA$ with our sample of red clump stars. On top of this, NLTE effects could be important in this range of metallicity for Ba II lines \citep{Mashonkina+1999,DOrazi+2009}.

Here we see clear decreasing trends of all $s$-process elements (except Pr), with a sort of plateau at young ages. Thus, our results go in the same line of the previous studies by \citet{Magrini+2018,Maiorca+2011}, but with more than double the number of clusters, and overall lower uncertainties in the measured abundances. We find that Ba is the element which reaches a larger value in this plateau, $\sim0.4$ dex. Lower values are found for Ce, Y, Nd ($\sim0.15$ dex), and Mo, Zr ($\sim0$ dex). Super-solar values of these abundances are also found by the very recent work of \citet{Spina+2020} which use the GALAH DR3 survey data. However, no clear trends with age are seen in their sample probably due to the smaller number of clusters with measured neutron-capture elements.

The spatial coverage of our sample goes out of the Solar neighbourhood with a range of $6<\rgc<11$ kpc (see Fig.~\ref{fig:XY}). In Fig~\ref{fig:XFe_age} we distinguish the clusters inside 1 kpc bubble around the Sun (orange), and those outside (blue). Both subsets have a similar age and metallicity distributions, though the $>1$ kpc sample is dominated by clusters in the outer disk, which also tend to be more scattered in the Z coordinate (Fig.~\ref{fig:XY}). 
\citet{Magrini+2018} also did a spatial analysis, comparing Solar neighbourhood vs inner disk clusters, concluding that focusing in the young age range ($<1-2$ Gyr), their Solar neighbourhood OCs ($6.5 < \rgc < 9.5$) have higher [X/Fe] than those in the inner disk ($\rgc<6$). We do not observe this behaviour, we see that the trends of the $s$-process elements are qualitatively satisfied in all distance regimes. However, our distance range is more limited than that of \citet{Magrini+2018} in the inner disk.

\section{Chemical clocks}\label{sec:chemclocks}
Several combinations of elements produced by different stellar progenitors, and thus produced at different timescales, have proven to be useful to derive empirical ages \citep[see e.g. discussion in][]{Nissen2015}. This is already seen in Fig.~\ref{fig:XFe_age}, where some [X/Fe] have steep trends with age, and others are flat \citep[see also extended discussion by][]{Delgado-Mena+2019}. 

\citet{Nissen2015} and \citet{Spina+2018}, among several other authors, analyzed the tight linear dependence of [Y/Mg] and [Y/Al] as a function of age with solar twins. Both are promising chemical clocks, presumably useful to retrieve ages up to a precision of $\sim0.5$ Gyr, in a limited range of metallicity \citep{Feltzing+2017}.
This is something expected taking into account the timescale of formation of those elements since nearly all odd-Z and $\alpha$ elements are synthesized by core-collapse supernovae \citep{Woosley+2002,Kobayashi+2020}, contrary to most $s$-process elements whose production is dominated by intermediate-mass stars in their AGB phase.

Some further studies \citep{Jofre+2020,Delgado-Mena+2019} found other possibilities of combinations of elements useful as chemical clocks candidates. A linear regression as a function of stellar age is able to retrieve some abundance ratios with slopes as significant as that of [Y/Mg] determined by \cite{Nissen2015}, for example. However, as pointed out by \citet{Delgado-Mena+2019} some of these ratios might hold only for certain ranges of metallicity. 

\subsection{The Solar neighbourhood}\label{sec:solarneighb}

\begin{figure*}
\centering
\includegraphics[width=\textwidth]{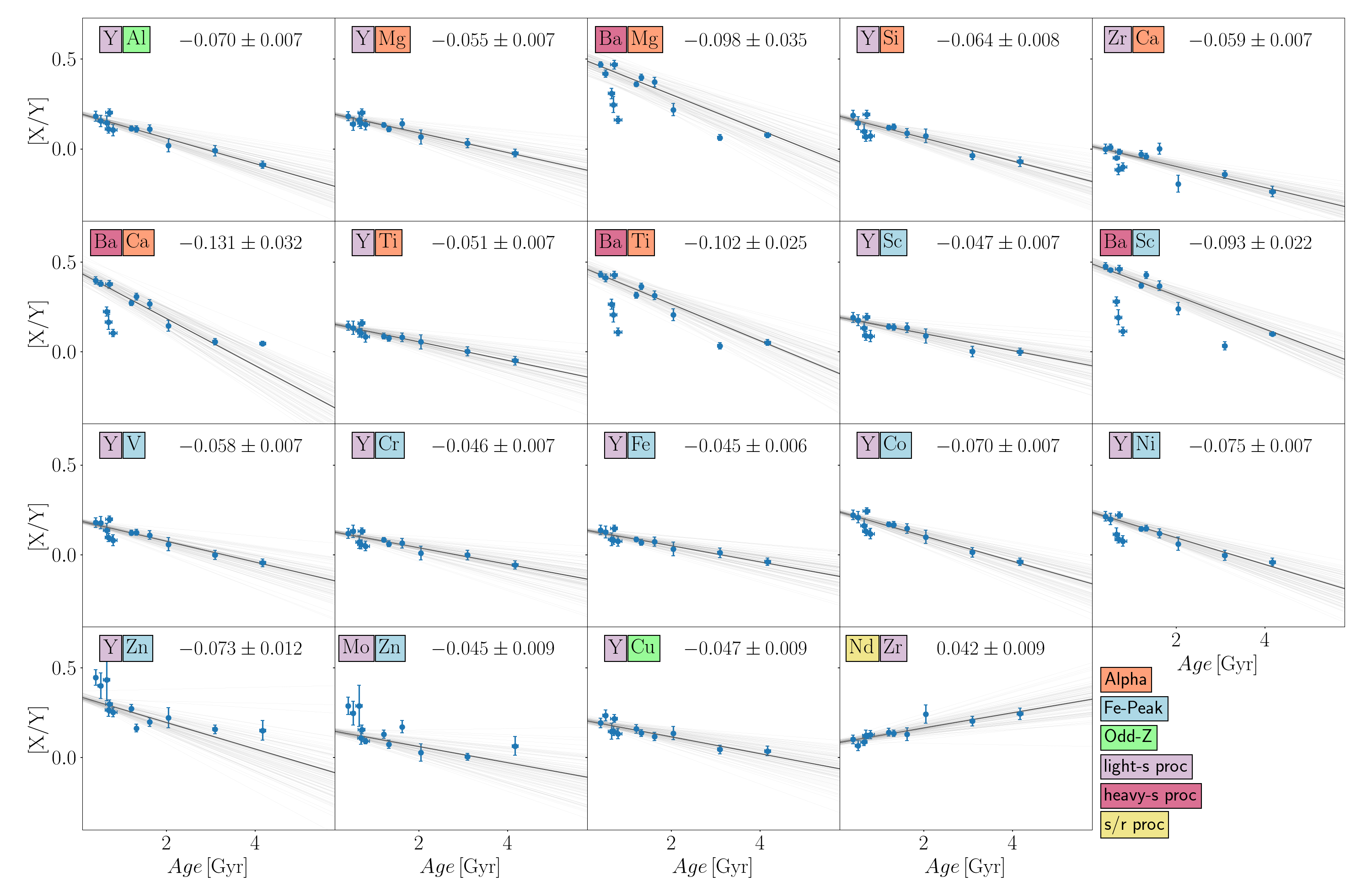}
\caption{Abundance ratios for which a significant level of correlation is found. Only clusters closer than 1 kpc and with more than one star are plotted. Errorbars in X axis are usually smaller than the point size. The best fit model is overplotted in black (the resulting slope written in each panel in $\mathrm{dex\,Gyr}^{-1}$), and the light gray lines correspond to the sampling of the MCMC. Element names are coloured depending on the chemical family where they belong to.}\label{fig:sign_corr}
\end{figure*}

In this subsection we use our sample of clusters to investigate all possible combinations of elements that show a significant correlation with age. We perform our analysis and plots as a function of age in a linear scale (instead of logarithmic scale as in Sect.~\ref{sec:XFe_age}) to be able to quantitatively compare our linear fits to those performed in the literature. We only use the subsample of clusters inside the 1 kpc bubble to reproduce similar conditions as in previous studies that investigate chemical clocks. None of the five flagged clusters in Sect.~\ref{sec:FinalSample} is in this subsample.

In general, we expect a trend when the timescale of production of the two elements involved differ (in particular, when the slopes of the trend found for the two elements as a function of age in Fig.~\ref{fig:XFe_age} go in opposite directions).
We have used the Bayesian outlier rejection algorithm explained in detail in \citet{Hogg+2010}\footnote{Our code is adapted from the one available in \url{https://www.astroml.org/book_figures/chapter8/fig_outlier_rejection.html}, which uses the python library PyMC3.} to perform a linear fit with uncertainties, to the different combinations of abundances. This algorithm is suitable for our case because our data is scarce, and helps to retrieve more robust estimates and uncertainties of the fitted parameters. The algorithm infers at the same time the two parameters of the fit together with a distribution of outliers. The model is run 15,000 times through a Markov Chain Monte Carlo (MCMC), and the best values of the slope and intercept are taken as the maximum of the posterior.
We have selected those for which we obtain slopes more significant than $3\sigma$, with absolute values $\geq 0.04\,\dexGyr$, and errors smaller than 0.03 $\dexGyr$. We find 19 significant combinations with our sample (including [Y/Mg] and [Y/Al]), plotted in Fig.~\ref{fig:sign_corr}.
In each panel we indicate the value of the slope with its uncertainty. In addition, following \cite{Jofre+2020}, we color-code each element by a nucleosynthetic family, to guide the eye and help in the interpretation.

Similarly to \cite{Jofre+2020}, we find that all selected ratios involve at least one element produced partially by the $s$-process. Y, Zr and Ba have a large contribution from the $s$-process \citep[more than $\sim$60\%][]{Bisterzo+2014}, because are mainly produced by intermediate-mass stars in their AGB phase, thus ejected to enrich the interstellar medium in an overall large timescale. The $r$-process, on the contrary, is believed to happen in core-collapse supernovae and possibly other more exotic places like neutron star mergers \citep{Thielemann+2017}, where a large density of neutrons is available. Nd has a significant contribution from the $r$-process.

We find that the "purest" $s$-process elements (Y, Zr and Ba) produce significant correlations when combined with $\alpha$-elements, remarkably with Mg, Si, Ca and Ti. Only Y has a significant correlation with Al. In these combinations Ba is the element for which we find the steepest slopes, which are always close to -0.1 $\dexGyr$. We recall the reader that Ba abundances can be affected by activity and NLTE effects as discussed in Sect.~\ref{sec:XFe_age}.
Y and Zr typically show smaller slopes, except for the combination with Al (and some Fe-peak elements) which gives -0.07 $\dexGyr$. The different behaviour of $s$-process elements, may reflect how differently their production evolves with time.
In general, our slopes have higher values with respect to \cite{Jofre+2020}, especially those involving Ba since our values are close to $-0.10\,\dexGyr$, whereas \citet{Jofre+2020} obtain values around $-0.04\,\dexGyr$. This is a reflection of the steeper [Ba/Fe] trend in Fig.~\ref{fig:XFe_age} seen in clusters and also in other studies (discussed in detail in Sect.~\ref{sec:XFe_age}), in comparison with field stars. We note that in the Ba relations there are persistently three outliers which are usually detected by our algorithm: the Hyades cluster, NGC~2632 and NGC~6997.

Nd is a heavy neutron-capture element with a significant contribution from both the $s$-process and the $r$-process, around 50\% of each path. 
We find positive slopes of Nd, when combined with the $s$-process element Zr (also produces a similar correlation with Ba but it does not get selected because of low significance). The sign of the slope goes in the opposite direction indicating that its production timescale is much quicker than that of Zr. This is also related with the different slope of the [Nd/Fe] vs age, in comparison with [Zr/Fe] seen in Fig~\ref{fig:XFe_age}.

We also find that Y has trends with almost all Fe-peak elements. Again, this is explained in Fig~\ref{fig:XFe_age}: a steep trend from Y which produces a slope when combined with the flat trend of Fe-peak elements. From the other $s$-process elements, only Ba produces a trend with Sc, and Mo with Zn. All Fe-peak elements have significant contributions from both core-collapse supernovae and type Ia supernovae, with slightly different yields depending on the element, except for Cu and Zn which are almost entirely produced by core-collapse supernovae, but with yields depending on metallicity. That makes it difficult to interpret our obtained trends in this family of elements. This result is not new, since \citet{Jofre+2020} also found some significant trends with the iron-peak element family, particularly Ni, Mn and Co, when combined with $s$-process elements.

We want to highlight that some of the trends can be very sensitive to the non uniform distribution in age of our (and any) sample of clusters, biased towards young ages, and thus very dependent of the abundances of the few old clusters. Even though our sample is the largest of clusters used to study age-abundance relations, some of our trends carry large uncertainties due to the low number of clusters. Additionally, on top of pure nucleosynthetic arguments, the trends have to be also understood in terms of the high complexity in the chemical substructures in the local chemical abundance space, as shown for example by \citet{Anders+2018}. In the same spatial volume, one can find several subgroups of stars belonging to different populations, including inner disk and outer disk migrators, stars in eccentric orbits, etc. This effect gets more important when the volume analyzed increases, like in the present work compared to the previous samples of solar analogues. Thus, the difference in the found slopes, and in the scatter of abundance at a given age, w.r.t. studies like \citet{Jofre+2020} have to be understood in the context of a chemo-dynamical picture of the Galaxy. This is discussed in the next section.

\subsubsection*{[Y/Mg] and [Y/Al]}

\begin{figure*}
\centering
\includegraphics[width=\textwidth]{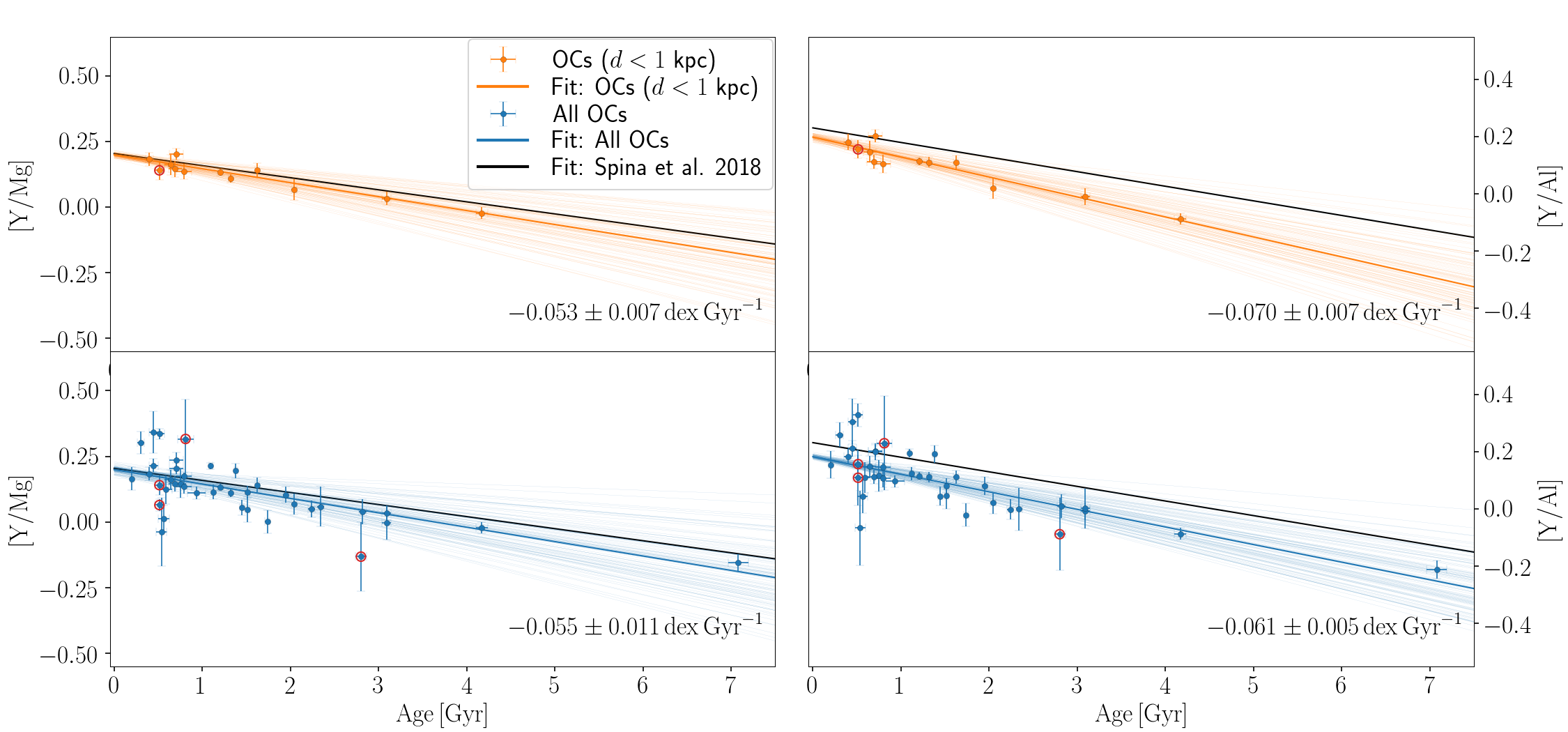}
\caption{[Y/Mg] (left) and [Y/Al] (right) abundances of our sample of clusters (with more than one measured star) as a function of age. In the top panel we plot the clusters closer than 1 kpc (orange), and in the bottom panel clusters at all distances. The clusters marked with red circles are those flagged in Sect.~\ref{sec:FinalSample}. The linear fits done for the close (all) clusters is plotted in orange (blue) lines, accounting for uncertainties. The values of the obtained slopes are indicated. The results of the linear fit performed by \citet{Spina+2018} are shown in black, for comparison purposes.}\label{fig:YMgAl}
\end{figure*}

The two most studied chemical clocks in the literature are [Y/Mg] and [Y/Al]. In Fig.~\ref{fig:sign_corr}, we have found that they exhibit significant trends with age when we use the sample of clusters in the local bubble. This shows that red giants also verify the age dependence of these two ratios as seen in the literature with Solar type stars. Here we aim to quantitatively compare our results with the trends traced by Solar twins for these two ratios of abundances \citep{Nissen2015,Spina+2018}.

We plot in the top panels of Fig.~\ref{fig:YMgAl} the obtained abundances of [Y/Mg] (left) and [Y/Al] (right) vs age of the clusters closer than 1 kpc, which have at least two members with measured abundances, which we consider to have more reliable uncertainties (coming from the dispersion of the star-by-star abundances). This is the same sample as the one used in Fig.~\ref{fig:sign_corr}, and the corresponding linear fit obtained:

\begin{equation}\label{eq:res_fit_c}
\begin{split}
\mathrm{[Y/Mg]} = 0.198\,(\pm0.014) - 0.055\,(\pm0.007)\cdot \mathrm{Age} \\
\mathrm{[Y/Al]} = 0.200\,(\pm0.013) - 0.070\,(\pm0.007)\cdot \mathrm{Age}
\end{split}
\end{equation}

The linear fit of the decreasing trend with age obtained by \cite{Spina+2018} is overplotted in black which corresponds to the slopes of $-0.046\pm0.002\,\dexGyr$ for [Y/Mg] and $-0.051\pm0.002\,\dexGyr$ for [Y/Al]. These values of the slopes are similar also to other studies \citet[e.g.][]{Nissen2015,TucciMaia+2016}. We find similar slopes as those obtained in the literature, though with larger uncertainties compared to the analysis done with Solar twins. This is probably due to the smaller number of clusters. We plot in red the only flagged cluster in Sect~\ref{sec:FinalSample} closer than 1 kpc and with more than 1 star which is NGC~6181. The best fit result does not change if we do not take into account this cluster.

The [Y/Al] slope is only compatible at $\sim2\sigma$ with that of \cite{Spina+2018}, and additionally, we find a zero point of $\sim$0.05 dex in the [Y/Al] abundance compared to them. Some studies using stars of different spectral types in clusters \citep[e.g.][]{Souto+2018} show that diffusion affects abundances of some elements such as Al of main-sequence stars, with differences as high as 0.1 dex. Later in stellar evolution, in the RGB, the convective envelopes tend to restore initial chemical abundances. Thus, it can be that diffusion is affecting abundances of the studies in the literature. The offset could also be explained if Al suffers from NLTE effects in giants \citep{Andrievsky+2008}, which we do not take into account in this work. It could also be related to the fact that some light elements can be affected by internal mixing processes in giants, which alter the relations between age and abundance, and can be significant for Al \citep{Smiljanic+2016}. This issue needs further investigation. This would mean that the [Y/Al] relations obtained with dwarfs would give biased results if one tried to apply it to date giants, or vice-versa.

\subsection{Beyond the local bubble}\label{sec:beyond}
Here, we want to investigate the spatial dependence of chemical clocks. And particularly, if the inclusion of distant OCs significantly changes the local derived slopes.

In the bottom panels of Fig.~\ref{fig:YMgAl} the obtained abundances of [Y/Mg] (left) and [Y/Al] (right) vs age of all clusters analysed in this paper are shown. The corresponding linear fit is also plotted, performed in the same way as for the clusters in the Solar neighbourhood.
We notice a remarkable larger dispersion when compared to the results of the local bubble, and to the studies of Solar twins in the literature. The results of these fits are:

\begin{equation}\label{eq:res_fit_f}
\begin{split}
\mathrm{[Y/Mg]} = 0.199\,(\pm0.017) -0.055\,(\pm0.011)\cdot \mathrm{Age}\\
\mathrm{[Y/Al]} = 0.182\,(\pm0.012) - 0.061\,(\pm0.005)\cdot \mathrm{Age}
\end{split}
\end{equation}

The two sets of slopes (Eq.~\ref{eq:res_fit_c} and ~\ref{eq:res_fit_f}) are compatible within $1\sigma$, though in [Y/Al] we obtain slightly flatter slopes when using the full sample compared to the close one. This last conclusion needs to be taken with caution, because the samples are small in numbers, and our fits carry large uncertainties. What we do obtain though in Fig~\ref{fig:YMgAl}, is that there appears a significantly larger scatter in abundances of the distant clusters compared to close ones. Thus, we interpret this as a hint that the abundance-age relations may be dependent on the spatial volume analyzed. Assuming that possibly the samples of stars get mixed due to dynamical processes of the Milky Way, then enlarging the analyzed volume introduces more scatter in these relations. Doing the fit with or without the flagged clusters (marked in red in the figure, see Sect.~\ref{sec:FinalSample}) retrieves very similar slopes: $-0.054\pm0.007$ and $-0.064\pm0.005\,\dexGyr$, for [Y/Mg] and [Y/Al].

\begin{figure*}
\centering
\includegraphics[width=\textwidth]{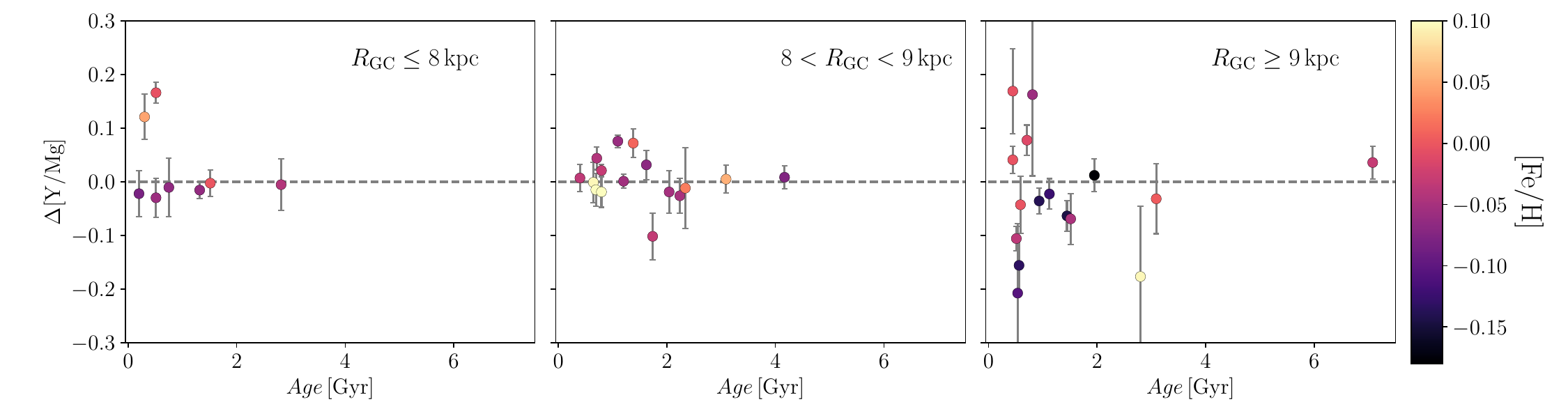}
\caption{Residuals in the [Y/Mg] (w.r.t. the linear fit for the closest sample Equation \ref{eq:res_fit_c}) vs age in three bins of Galactocentric radius indicated in each panel. The color code is the derived [Fe/H] abundance.}\label{fig:compareRgc}
\end{figure*}

We plot in Fig~\ref{fig:compareRgc} the residuals\footnote{w.r.t. the linear fit done for the sample of close clusters: Equation \ref{eq:res_fit_c}} of the [Y/Mg] vs age of the sample of clusters separated in bins of $\rgc$: inner disk ($\rgc\leq8$ kpc), Solar neighbourhood ($8<\rgc<9$ kpc) and outer disk ($\rgc\geq9$ kpc).
By doing a weighted standard deviation of the residuals in these three ranges, we obtain larger dispersions in the inner and outer disk of 0.07 and 0.14 dex, respectively, in comparison with the Solar radius which is 0.04 dex.
The most discrepant clusters tend to be young ($\sim500$ Myr) in both the inner and outer samples.
These clusters also tend to have large uncertainties in our [Y/Mg] abundances ($\sim0.1$ dex). However, we notice in the plot that there are cases where the uncertainties in the [Y/Mg] determination are small, but provide large residuals. This is the case of the cluster NGC~6705 (M~11, $Age=300$ Myr) which gives a difference in abundance of 0.12 dex with an uncertainty of $\delta_{\mathrm{[Y/Mg]}}=0.04$ based on 12 members. This cluster is nevertheless peculiar for its $\alpha$-enhancement \citep{Casamiquela+2018}.

We do not see a clear dependence of the residuals as a function of the metallicity or the height above the Galactic plane. However, we notice that clusters in the outer disk tend to be more scattered vertically (as can be seen in Fig.~\ref{fig:XY}).

Translating abundance scatter into age scatter, we obtain that the [Y/Mg] relations, even inside the $\sim$1 kpc bubble around the Sun, could predict ages with a precision of $\sim$1 Gyr, at best.
This result goes in the same line as the recent work by \citet{Morel+2020}, who investigated several age-abundance relations using 13 stars in the Kepler field with asteroseismic ages. They find that the seismic and abundance-based ages differ on average by $\sim$1.5 Gyr. They attribute this to a variety of causes including the presence of small zero-point offsets between chemical abundances and those used to construct the age-abundance calibrations.

A similar experiment was done in the recent work by \citet{Casali+2020}. They noticed that ages computed from [Y/Mg] of a subsample of OCs towards the inner disk ($\rgc<7$ kpc) were in general overestimated with respect to literature (isochrone) ages. This sample of clusters in the inner Galaxy is, in general, more metal-rich than the rest of their sample ($\mathrm{[Fe/H]}\gtrsim0.2$ dex), and the differences they obtain in age are between 2-8 Gyr. They attribute these large discrepancies to the fact that the yields of $s$-process elements have non-monotonic dependencies with metallicity. This metallicity dependence is particularly important for heavy elements, which are secondary elements produced depending on the quantity of seeds available (iron-peak elements). In our sample we do not see a striking dependence in the residuals as a function of metallicity. This could be because our metallicity range is a bit narrower than theirs, and particularly we have less metal-rich clusters.

We have eight clusters in the inner $\rgc$ range ($\rgc\leq8$ kpc), and the clusters NGC~6645 and NGC~6705 are the discrepant ones. Otherwise, the scatter of the residuals of the other clusters is 0.01 dex. On the contrary, we observe a larger offset and dispersion towards the outer Galaxy ($\rgc\geq9$ kpc), based on 16 clusters. Some studies have argued that OCs suffer radial mixing processes in the same degree as single field stars \citep{Casamiquela+2017}. Therefore, an interpretation of the large scatter in the abundances of clusters is that going out of the very local Solar neighbourhood traced by solar twins (typically confined within $\sim$100 pc around the Sun), there is a higher chance to include in our sample stars/clusters which have migrated radially from their birthplaces hence tracing a different chemical enrichment history. Thus, simple abundance-age relations would not be valid anymore due to the complexity introduced by the dynamical processes.

\section{Conclusions}\label{sec:conclusions}
Clusters are good complements to solar twins to investigate trends of chemical abundance ratios with age. Their ages can be estimated with high precision through their color magnitude diagrams plus the information of \emph{Gaia} parallaxes, therefore they allow to probe outside the local bubble.

We have homogeneously analyzed high-resolution spectra from \NGoodstars\ red clump stars in \NClusters\ open clusters ranging ages between 140 Myr and 7 Gyr and metallicities between -0.2 to 0.2 dex. We derive atmospheric parameters with typical uncertainties of 18 K in $\teff$ and 0.045 dex in $\logg$. We retrieve abundances of \NElements\ chemical species using a line-by-line technique using spectral synthesis fitting for all stars with respect to a giant in the Hyades cluster. Then we transform to solar-scaled abundances using a solar analogue of the Hyades analyzed with respect to the Sun. This has proven to be a useful way to retrieve high-precision abundances of stars of different spectral types inside a cluster, without the need of observing solar twins, which are fainter. We retrieve cluster mean abundances with typical scatters of 0.02 dex in most of the elements, except for heavy elements, for which dispersions are larger.

We analyze the behaviour of [X/Fe] abundances with $\log Age$. Elements from Na to Zn exhibit generally flat trends, with large scatters in Mn, Cu and Zn, and remarkable zeropoints in Na, Mn and Zn. We find a knee in most neutron capture elements at around 1 Gyr, followed by a decreasing trend of abundance with $\log Age$, particularly in Ba, Ce, Y, Nd, Zr and Mo. The mentioned elements exhibit a plateau at young ages ($<1$ Gyr) with different values (Ba being the largest one at $\sim$0.4 dex).

Using the sample of clusters inside a 1 kpc bubble around the Sun, we find 19 different [X$_1$/X$_2$] combinations which have strong correlations with age in our metallicity range, particularly those which involve elements produced via the $s$-process. We find that Y, Zr and Ba (which have a large contribution to the $s$-process) always produce correlations when combined with $\alpha$-elements. Among all significant combinations, we find the largest slopes for those with Ba, such as [Ba/Mg] ($0.098\pm0.035\,\dexGyr$). However, we highlight that according to literature, Ba II lines can be affected by activity in young stars, and on top of this NLTE corrections may be needed. We compare some of the slopes with those derived by \citet{Jofre+2020}, and we find that in general our slopes are steeper. Special caution needs to be taken in this type of comparison since the age distribution of our (and any) sample of clusters is biased towards young ages, thus slopes are very sensitive to the values of old clusters. The discovery of new clusters in our Galaxy thanks to \emph{Gaia} is going to improve this situation and help future studies in this direction.

A linear fit to [Y/Mg] and [Y/Al] vs cluster ages retrieves similar slopes as those already published in the literature $-0.055\pm0.007\,\dexGyr$ and $-0.070\,\pm0.007\,\dexGyr$, respectively. Our [Y/Al] trend differs in slope w.r.t. other authors, and shows an overall offset in abundances which may be related to NLTE effects or mixing in giants, or alternatively due to diffusion in main sequence stars used for the literature studies. This needs to be further assessed before using this clock to date stars. The found differences also point towards some difficulties in computing ages using chemistry alone.

We investigate the validity of the abundance-age relations outside the local bubble. In [Y/Al] we obtain a slightly flatter slope when using the full sample of clusters, though this is not the case with [Y/Mg]. In both ratios we see a larger scatter introduced. We interpret this as a hint that the chemical clocks may not be as universal as thought, but instead they probably have a dependendence on the spatial volume analyzed.
The residuals of the [Y/Mg] fit have values around 0.04 dex for the clusters in the Solar radius ($8<\rgc<9$ kpc). For the outer Galaxy ($\rgc\geq9$ kpc) the comparison of abundances of the 16 clusters seem to disagree more, with a scatter of 0.14 dex in the residuals. We show that these differences are not fully explained by uncertainties in the [Y/Mg] value. This scatter translates into an age precision of 1 Gyr at best.

Thanks to the spatial coverage of our sample of clusters, our results point to a non-universality of the abundance-age relations, in the sense that expanding the spatial volume analyzed may increase the chance of including migrators which will certainly trace a different enrichment history. Thus, one needs to understand the abundance trends also in terms of the complexity of the chemical space introduced by the Galactic dynamics, on top of pure nucleosynthetic arguments \citep[e.g.][]{Anders+2018}. Particular caution needs to be taken when trying to apply these relations to date samples of stars out of the local bubble.

\begin{acknowledgements}
We thank the anonymous referee for the useful comments which helped to improve the quality and results of the paper.
This work has made use of data from the European Space Agency (ESA) mission \emph{Gaia} (\url{http://www.cosmos.esa.int/gaia}), processed by the \emph{Gaia} Data Processing and Analysis Consortium (DPAC, \url{http://www.cosmos.esa.int/web/gaia/dpac/consortium}). We acknowledge the \emph{Gaia} Project Scientist Support Team and the \emph{Gaia} DPAC. Funding for the DPAC has been provided by national institutions, in particular, the institutions participating in the \emph{Gaia} Multilateral Agreement.
This research made extensive use of the SIMBAD database, and the VizieR catalogue access tool operated at the CDS, Strasbourg, France, and of NASA Astrophysics Data System Bibliographic Services. We also thank the staff maintaining the public archives of ready-to-use spectra at ESO, NOT, OHP, CFHT and TBL.

This work has used data from the ESO programs: 091.C-0438(A), 079.C-0131(A), 075.C-0140(A), 077.C-0364(E),076.C-0429(A), 0101.C-0274(A), 0104.C-0358(A),0102.C-0812(A), 092.C-0282(A), 076.B-0055(A), 60.A-9036(A),094.C-0297(A), 095.C-0367(A), 383.C-0170(A), 080.C-0071(A),081.C-0119(A), 082.C-0333(A), 083.C-0413(A), 079.C-0329(A),099.C-0304(A), 0100.C-0888(A), 266.D-5655(A),185.D-0056(C), 193.B-0936(D), 185.D-0056(I), 70.D-0421(A),088.C-0513(B), 094.D-0596(A), 086.C-0145(A), 380.C-0083(A), 097.A-9022(A), 085.D-0093(A), 072.C-0393(D),0100.A-9018(A), 083.A-9011(A)

L.C., C.S., Y.T. and N.L. acknowledge support from "programme national de physique stellaire" (PNPS) and from the "programme national cosmologie et galaxies" (PNCG) of CNRS/INSU. We acknowledge the support of the scientific cooperation program ECOS-ANID Number 180049. L.C. acknowledges the support of the postdoc fellowship from French Centre National d’Etudes Spatiales (CNES). P.J. acknowledges financial support of FONDECYT Iniciaci\'on grant Number 11170174 and FONDECYT Regular grant Number 1200703. 
This work was partially supported by the Spanish Ministry of Science, Innovation and University (MICIU/FEDER, UE) through grant RTI2018-095076-B-C21, and the Institute of Cosmos Sciences University of Barcelona (ICCUB, Unidad de Excelencia ’Mar\'{\i}a de Maeztu’) through grant CEX2019-000918-M.

\end{acknowledgements}

\bibliographystyle{aa} 
\bibliography{biblio2_v1,biblio_v4}

\begin{appendix}
\section{Additional figures and tables}

\begin{table*}[h!]
\caption{\label{tab:stars}Details of the spectra of the observed stars identified by their \emph{Gaia} DR2 source id. The instrument, origin (own observations or archival data) and signal-to-noise of each spectrum is indicated. Only a portion of the table is shown here, the complete version will be available online.}
\setlength\tabcolsep{3.5pt}
\centering
\begin{tabular}{lrrccccccccc}
 \hline
 Cluster & Gaia DR2 sourceid & Instrument & Origin & S/N  \\
 \hline
 IC 4756 & 4283939920251942144 & FIES   &Observations &  81\\
 IC 4756 & 4283939920251942144 & HARPS  &Archive      &  66\\
 IC 4756 & 4283939920251942144 & HERMES &Observations &  81\\
 IC 4756 & 4283939920251942144 & UVES   &Archive      & 114\\
 IC 4756 & 4283940671842998272 & HARPS  &Archive      &  65\\
 IC 4756 & 4283940671842998272 & HERMES &Observations &  82\\
 ... & & & & \\
 \hline
\end{tabular}
\end{table*}

\begin{table*}[h!]
\caption{\label{tab:clusters}Properties of the 47 OCs studied in this work. We indicate the distance $d$, Galactocentric radius $\rgc$, height above the plan $Z$ and used ages. The mean abundances and their errors (computed as explained in the main text), and the number of stars used in each cluster are in parenthesis for all chemical elements analyzed. Only a portion of the table is shown here, the complete version will be available online.}
\setlength\tabcolsep{3.5pt}
\centering
\begin{tabular}{lrrccccccccc}
 \hline
 Cluster & $d$ [pc] & $\rgc$ [pc] & $Z$ [pc] & $Age\,\mathrm{[Gyr]}$ & [\ion{Fe}{I}/H] & [\ion{Fe}{II}/H] & [\ion{Na}{I}/H] & [\ion{Al}{I}/H] & ...  \\
 \hline
 Hyades   & 47   & 8384 & -17 & $0.79\pm0.09$ & $ 0.14\pm 0.02$  (3) & $0.161\pm0.007$ (3) & $ 0.12\pm0.01$ (3) & $ 0.10\pm0.02$   (3) & ...  \\
 NGC 6991 & 569	 & 8333 & 15  &	$1.62\pm0.03$ &	$-0.030\pm0.002$ (4) & $ 0.06\pm0.03$  (4) & $-0.24\pm0.02$ (4) & $-0.014\pm0.006$ (4) & ... \\
 NGC 6811 & 1128 & 8205 & 234 &	$1.09\pm0.01$ &	$-0.007\pm0.002$ (6) & $0.098\pm0.008$ (6) & $-0.22\pm0.01$ (6) & $0.012\pm0.009$  (6) & ... \\
 ... & & & & & & & & & \\
 \hline
\end{tabular}
\end{table*}

\begin{figure*}[h!]
\centering
\includegraphics[width=0.7\textwidth]{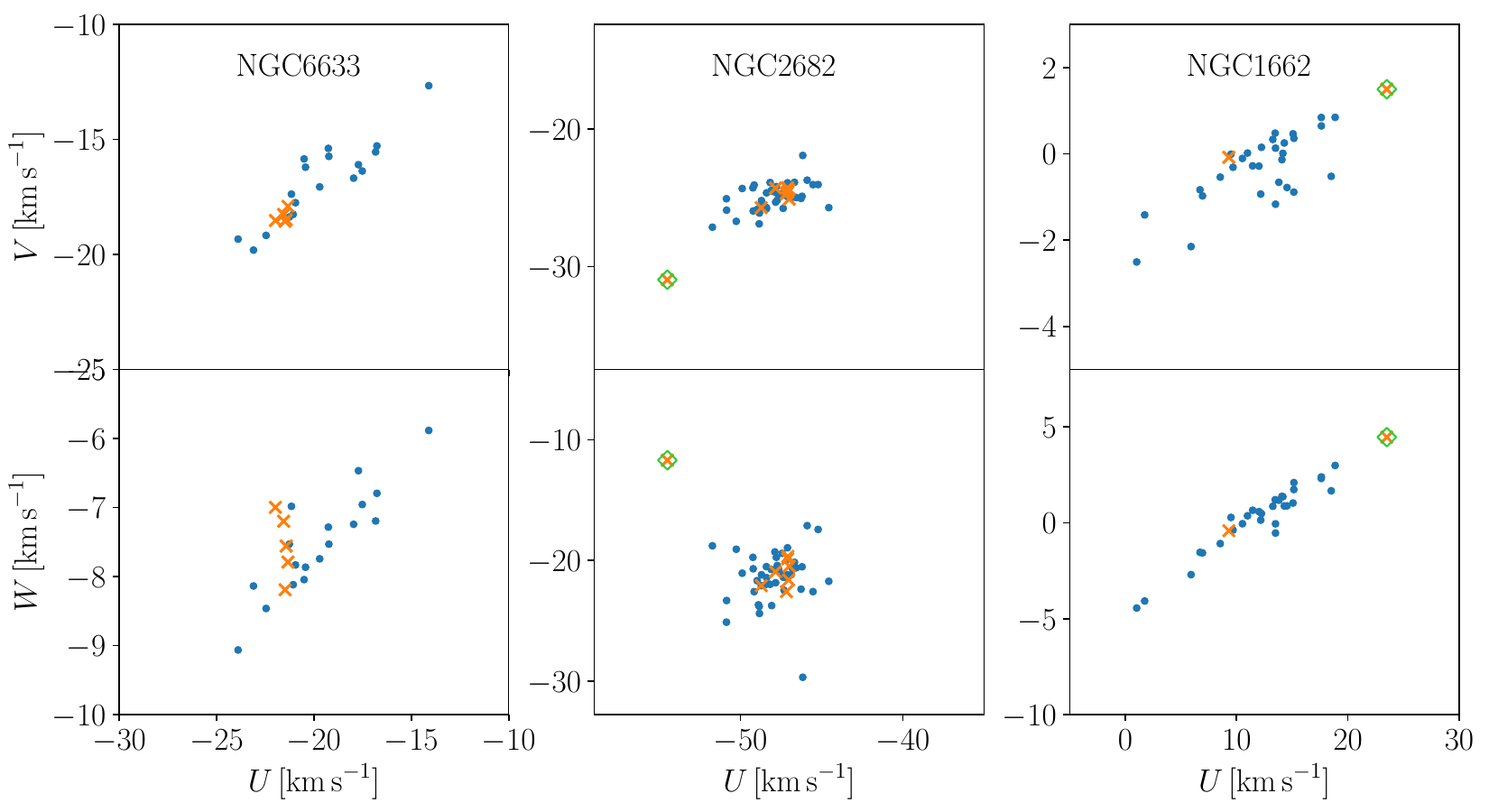}
\caption{Example of $(U,V,W)$ distribution of the member stars (blue) of the clusters NGC~6633, NGC~6281 and NGC~1662. Stars included in the spectroscopic analysis in this work are the orange crosses, and the stars discarded by discrepant kinematics are marked in green rhombs. The sample of member stars are those in \citet{Cantat-gaudin+2018} with membership probability $>0.7$, which have radial velocity determination in \emph{Gaia} DR2. Not all plots have the same scale.}\label{fig:exampleUVW}
\end{figure*}

\begin{figure*}[h!]
\centering
\includegraphics[width=0.7\textwidth]{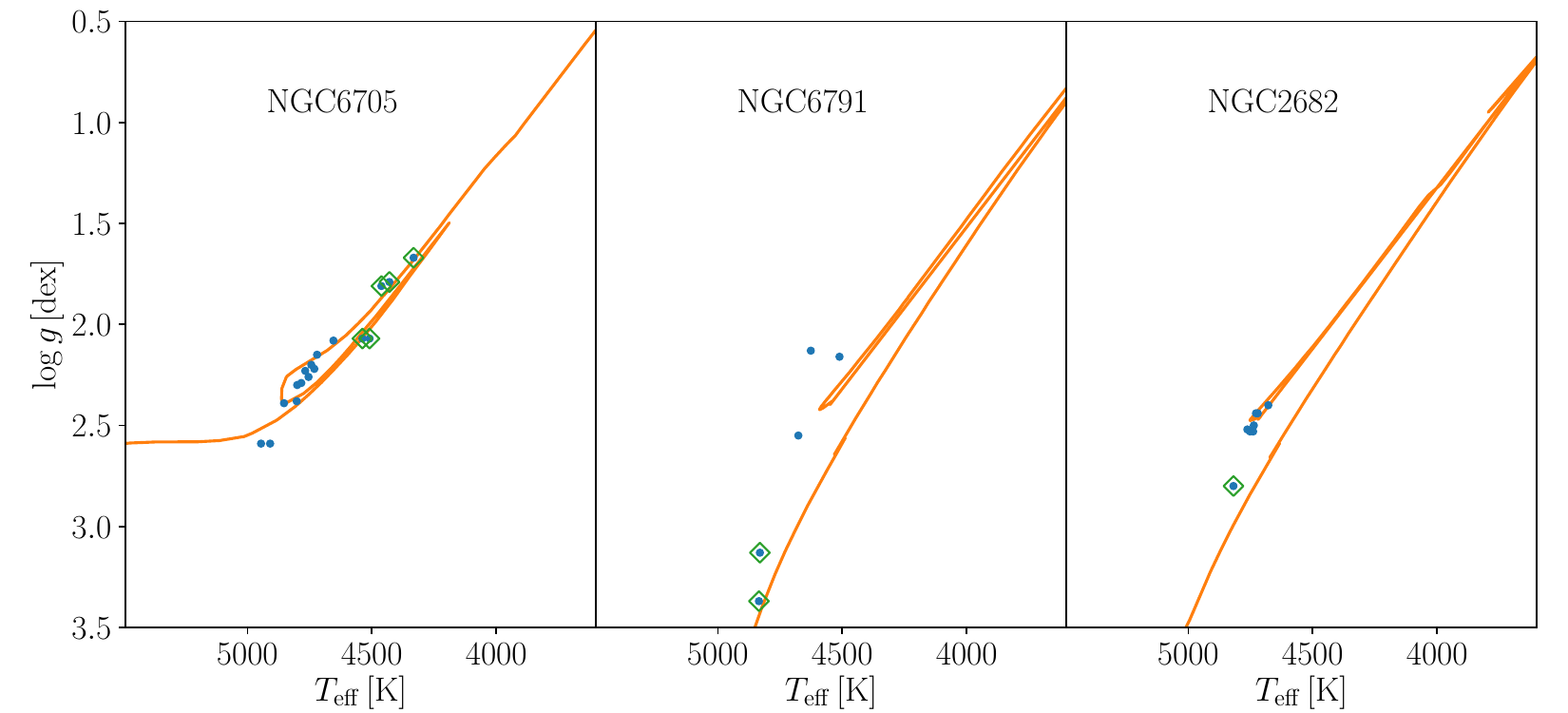}
\caption{Example $\teff-\logg$ distribution of the spectroscopic targets (blue) of the clusters NGC~6705, NGC~6791 and NGC~2682. Isochrones of the corresponding age and metallicity of the cluster are overplotted, and the members discarded for being unlikely red clump stars are marked in green rhombs.}\label{fig:exampleHR}
\end{figure*}

\begin{figure*}[h!]
\centering
\includegraphics[width=\textwidth]{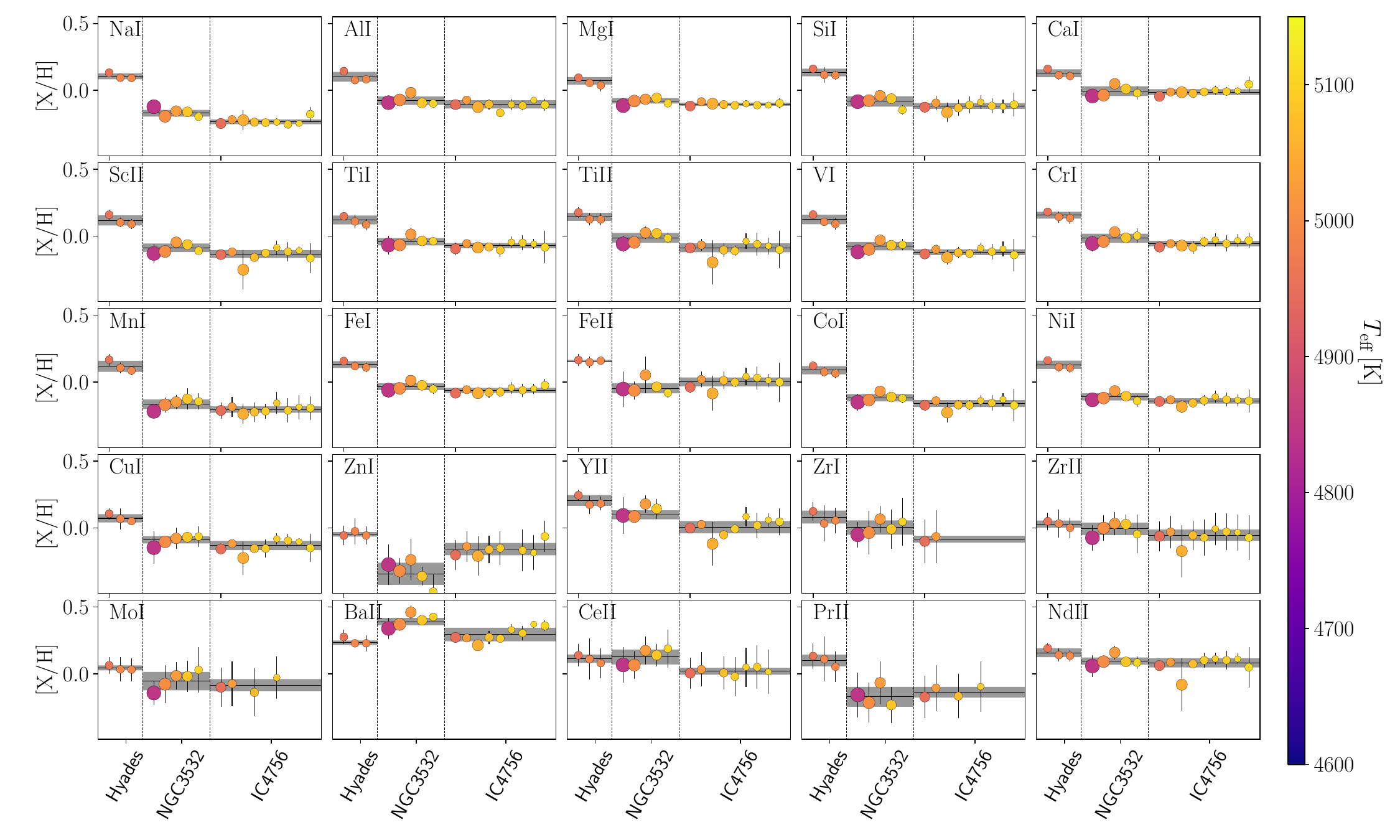}
\caption{[X/H] solar-scaled abundances of all measured chemical species for all stars in the Hyades, NGC~3532 and IC~4756. Stars are coloured and sorted by increasing $\teff$, and the size of the points is proportional to the $\logg$. The solid black line indicates the mean abundance, and the coloured band the standard deviation. For several elements (Zr, Ce, Pr) it was not possible to retrieve reliable values of abundance for the hottest stars.}\label{fig:NGC3532}
\end{figure*}

\begin{figure*}[h!]
\centering
\includegraphics[width=\textwidth]{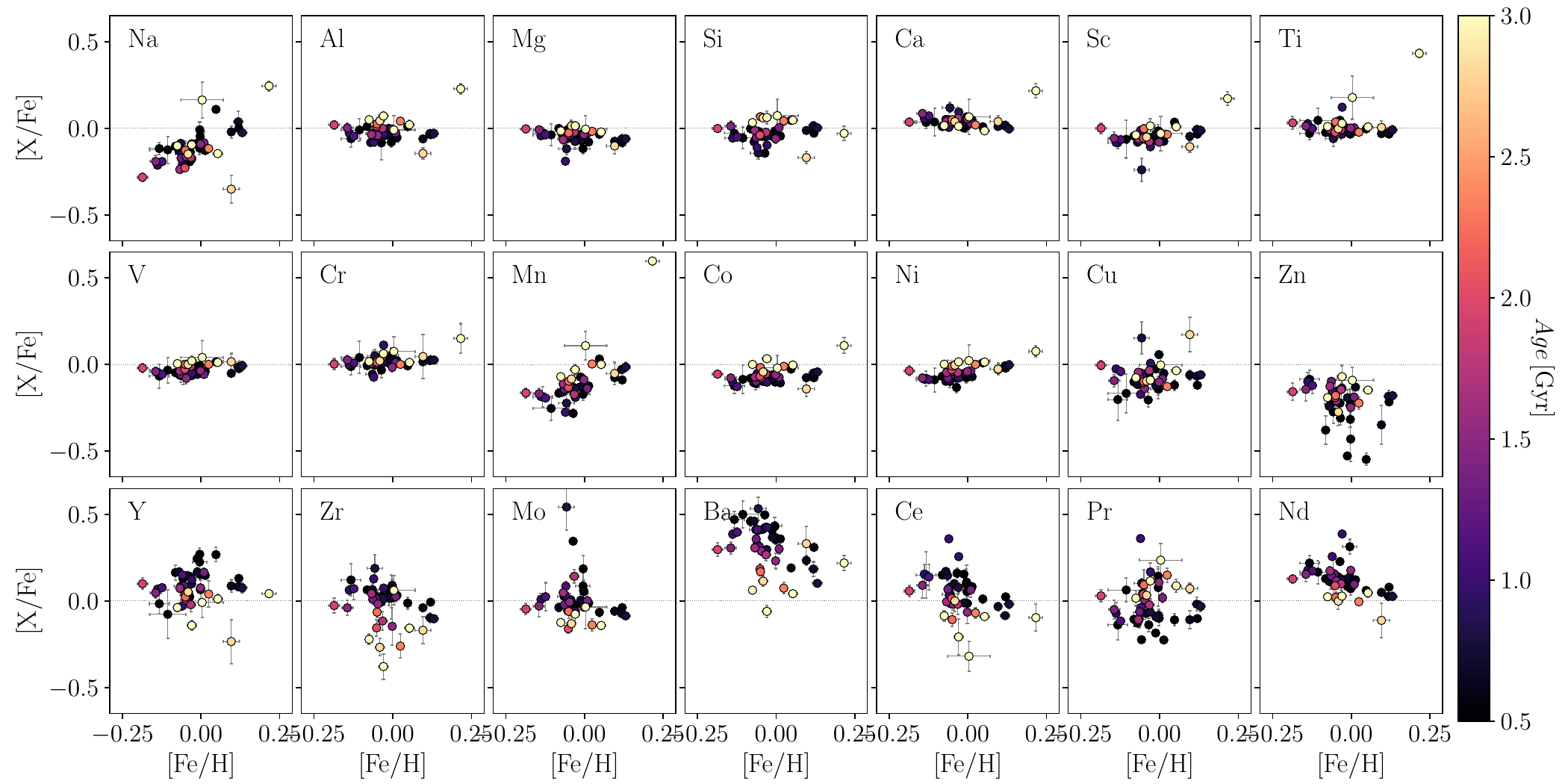}
\caption{[X/Fe] abundances of all measured elements as a function of [Fe/H] coloured by cluster age.}\label{fig:XFe_FeH}
\end{figure*}

\end{appendix}

\end{document}